\documentclass[prb,aps,twocolumn,floats,superscriptaddress,showpacs]{revtex4}
\usepackage{graphicx}
\renewcommand{\vec}[1]{{\bf #1}}

\newcommand{\nin}{\noindent}

\newcommand{\be}{\begin{equation}}
\newcommand{\ee}{\end{equation}}
\newcommand{\bea}{\begin{eqnarray}}
\newcommand{\eea}{\end{eqnarray}}
\newcommand{\lb}{\left[}
\newcommand{\rb}{\right]}
\newcommand{\lp}{\left(}
\newcommand{\rp}{\right)}
\newcommand{\lf}{\left\{}
\newcommand{\rf}{\right\}}
\newcommand{\la}{\left <}
\newcommand{\ra}{\right >}
\newcommand{\tiafm}{$\triangle$IAFM\,}
\renewcommand{\r}{{\bf r}}
\newcommand{\R}{{\bf R}}
\renewcommand{\P}{{\cal P}}
\renewcommand{\j}{{\bf j}}
\newcommand{\A}{{\cal A}}
\newcommand{\F}{{\cal F}}

\newcommand{\N}{{\cal N}}
\newcommand{\E}{\bf E}
\newcommand{\Vg}{V_{\rm g}}
\renewcommand{\k}{{\bf k}}

\begin{document}
\title{Correlated electron states and transport in triangular arrays}
\author{D.~S.~Novikov}
\email{dima@alum.mit.edu}
\affiliation{
Center for Materials Science \& Engineering,  Physics Department, \\
Massachusetts Institute of Technology, 
77 Massachusetts Avenue, Cambridge MA 02139}
\affiliation{
Department of Electrical Engineering and Department of Physics,
Princeton University, Princeton, NJ 08544}
\author{B.~Kozinsky}
\author{L.~S.~Levitov}
\affiliation{
Center for Materials Science \& Engineering,  Physics Department, \\
Massachusetts Institute of Technology, 
77 Massachusetts Avenue, Cambridge MA 02139}
\date{August 10, 2005}

\begin{abstract}

We study correlated electron states in frustrated geometry of a triangular lattice. 
The interplay of long range interactions and finite residual entropy 
of a classical system gives rise to unusual 
effects in equilibrium ordering as well as in transport.
A novel correlated fluid phase is identified in a wide range of densities 
and temperatures above freezing into commensurate solid phases.
The charge dynamics in the correlated phase is described in 
terms of a height field, its fluctuations, and topological defects.
We demonstrate that the height field fluctuations give rise 
to a ``free'' charge flow and finite dc conductivity.
We show that freezing into the solid phase,
controlled by the long range interactions, 
manifests itself in singularities of transport properties.
\end{abstract}


\maketitle

%%%%%%%%%%%%%%%%%%%%%%%%%%%%%%%%%%%%%%%%%%%%%%%%%%%%%%%%%%%%%%%%%%%%%%%%%%
\section{Introduction}
\label{sec:intro}

The properties of geometrically frustrated systems 
are complex due to the presence of an energy landscape with many 
degenerate or nearly degenerate minima.\cite{Ramirez}
These systems exhibit qualitatively new effects, such as
an extensive ground state degeneracy and
suppression of freezing down to zero temperature. 
The latter phenomena 
traditionally have been studied in the context
of antiferromagnetic spin models on appropriately chosen lattices,
and other models with short range interactions. 

Below we demonstrate that similar effects of geometrical
frustration can naturally arise in a classical {\it charge} system.
Namely, we consider ordering and dynamics of 
classical electrons on a triangular lattice with repulsive Coulomb 
interaction between charges on different lattice sites.
The frustration reveals itself in the phase diagram,
and also makes charge dynamics and transport very unusual.
Here we focus on the electric transport as a natural means 
to study ordering types and phase transitions in a charge system.

Two-dimensional frustrated lattices arise in a variety of experimental systems.
Artificial structures, 
such as Josephson junction arrays\cite{Josephson-arrays} 
and arrays of quantum dots,\cite{dot-arrays}
have recently become available.
One attractive feature of these systems
is the control of the Hamiltonian and,
in particular, of frustration,
by the system design. 
Also, experimental techniques available for 
probing magnetic flux or charge ordering, 
such as electric transport measurements and scanning probes,
are more diverse and flexible than those conventionally used 
to study magnetic or structural ordering in solids. 
There have been extensive theoretical
\cite{Josephson-arrays-theory,Josephson-arrays-theory-recent}
and experimental\cite{Josephson-arrays} studies of 
phase transitions and collective phenomena in Josephson arrays.  
Another example of frustrated charge systems is provided by novel 
superconducting materials based on CoO$_2$.\cite{highTc-tria}

In the present work we discuss 
a realization of a frustrated charge system 
in a 2d triangular array of quantum dots.
Recent progress in the epitaxial and lithographic techniques 
made it possible to produce\cite{dot-arrays} 
regular and irregular arrays of quantum dots, in which the size of an
individual dot can be tuned in the 10-100 nm range
with the rms size distribution of 10-20\%.
Such arrays are made of InAs or Ge islands
embedded into the semiconductor substrate. 
This stimulated experimental 
\cite{dot-arrays-experiment} and theoretical\cite{dot-arrays-theory} 
investigation of electronic ordering and transport in these arrays.
Capacitance and conductivity measurements 
\cite{dot-arrays-experiment} show the quantized nature of charging of
the dots. However, the interdot Coulomb interaction in such
systems is weak compared to the individual dot charging energy 
as well as potential fluctuations due to disorder in the substrate.

A promising system fabricated and studied recently\cite{Bawendi} involves
nanocrystallite quantum dots which are synthesized with high 
reproducibility, of diameters $\sim 1.5-15$ nm  
tunable during synthesis, with a narrow size 
distribution ($<5\%$ rms). These dots can be forced to assemble into
ordered 3d closely packed colloidal crystals,\cite{Bawendi}
with the structure of stacked 2d triangular lattices. 
High flexibility and structural control open a possibility
to study effects inaccessible in 
the more traditional self-assembled quantum dot arrays fabricated 
using epitaxial growth techniques. 
In particular, the high charging energy 
that can reach or exceed the room temperature scale, 
and the triangular lattice 
geometry of the dot arrays\cite{Bawendi} 
are of interest from the point of view of
exploring novel aspects of charge ordering and 
transport.\cite{qdots-photo,qdots-transport}

From a theoretical viewpoint, charge ordering is closely related to,
or can be interpreted in terms of, a suitably chosen spin system.
Spin models, which serve as a paradigm in a theory of critical phenomena, 
have applications to ordering in different systems,
such as the phase transitions in adsorbed monolayers.\cite{denNijs} 
In our analysis we map the charge problem 
onto the triangular antiferromagnetic Ising spin problem. %(\tiafm). 
In the situation of interest, when charging energy enforces single or zero
occupancy of the sites, one can interpret 
the occupied and unoccupied dots as an `up spin' and `down spin' states.
Since the like charges
repel, the corresponding effective 
spin interaction is indeed of an antiferromagnetic kind.
Appropriately, charge density plays the role of spin density, 
and the gate voltage corresponds
to an external magnetic field. 
Besides, one can map the offset charge disorder (random potentials on the dots)
onto a random magnetic field in the spin problem.

There are, however, a few theoretical and experimental aspects of
the charge-spin mapping that make the two problems not entirely equivalent.
First, charge conservation in the electron problem
gives rise to a constraint on total spin in the associated spin problem.
This leads to a dynamical constraint, namely blocking of single spin
flips. Microscopically, spin conservation requires the Kawasaki (or type B) 
dynamics\cite{Kawasaki} 
as opposed to the nonconserving Glauber (or type A) dynamics.\cite{Glauber} 
This makes no difference with regard to the 
thermodynamic state at equilibrium, since the system with fixed total spin 
is statistically equivalent to the grand
canonical ensemble. However, the order parameter conservation manifests itself 
both in a slower dynamics\cite{Hohenberg-Halperin} 
and in collective transport properties 
of the correlated fluid phase discussed below.

Another important difference between the charge and spin
problems is in the form of interaction. 
Spin systems are usually described by a nearest neighbor interaction.
In the spin problem relevant for this work, the triangular Ising
antiferromagnet (\tiafm) with the nearest neighbor interaction,
an exact solution in zero field has been obtained by
Wannier\cite{IsingAFM} who demonstrated that there is no ordering phase transition 
at any finite temperature. 
In contrast, in the charge problem studied in this article the long range 
Coulomb interaction makes the phase diagram more rich.  
We find phase transitions at finite temperature 
for certain charge densities.

One of the main objects in our focus is the {\it correlated fluid} phase,
that arises at relatively low temperatures due to geometric frustration 
preventing freezing.
In this phase, equilibrium fluctuations and transport exhibit strong 
correlations. We employ a nonlocal description in terms of the 
height field, used earlier in spin models,
\cite{Blote82,Blote84}
to map charge dynamics onto Gaussian fluctuations
of the height surface in the presence of topological defects (dislocations).
We find that both 
the height field fluctuations and the presence of defects
contribute to low-temperature transport properties.

The outline of the paper is as follows.
In Section~\ref{sec:themodel} a charge Hamiltonian is introduced 
and stochastic Monte Carlo (MC) dynamics 
is defined, based on the correspondence
between the charges and Ising spins.
This dynamics is used to obtain the phase diagram
(Section~\ref{sec:phase-diagram})
and to study the dc conductivity
as a function of temperature and electron density
(Section~\ref{sec:conductivity}). 

In Section~\ref{sec:height-variable}
a height field order parameter is defined and used 
to describe the charge ordering.
In Section~\ref{sec:cond-disl} we compare the contributions
to transport due to
the height surface fluctuations and the topological defects.
Subsequently, in Section \ref{sec:correlated-phase},
we study the height variable fluctuations and evaluate their effective stiffness
from the MC dynamics.
By comparing the stiffness to the universal value we rule out
the Berezinskii-Kosterlitz-Thouless transition and prove
that the defects are always unbound.  
In Section~\ref{sec:roughening} 
we present scaling arguments that support the conclusions of 
Section~\ref{sec:correlated-phase}, and set bounds on the height surface stiffness.
In Section~\ref{sec:dynamics} we consider dynamical fluctuations
of the height variable and estimate their effect on dc conductivity.

%%%%%%%%%%%%%%%%%%%%%%%%%%%%%%%%%%%%%%%%%%%%%%%%%%%%%%%%%%%%%%%%%%%%%%%%%%
\section{The Model}
\label{sec:themodel}

\nin
The Hamiltonian ${\cal H}_{\rm el}$
of a quantum dot array
describes charges
$q_i$ on the dots, their
Coulomb interaction
as well as coupling to the background disorder
potential $\phi(\vec r)$
and to the gate potential $V_{\rm g}$:
\begin{equation}\label{E_tot}
{\cal H}_{\rm el}=\frac{1}{2}\sum\limits_{i,j}V(\vec r_{ij})q_i q_j
+\sum\limits_{\vec r_i} (V_{\rm g}+\phi(\vec r_i))q_i \,.
\end{equation}
Here the
vectors $\vec r_i$ run over a triangular lattice with the 
lattice constant $a$,
and $\vec r_{ij}=\vec r_i-\vec r_j$. 
The interaction $V(\vec r)$ 
includes a term describing
screening by the gate:
\begin{equation}\label{interaction}
V(\vec r_{ij}\ne 0)=
\left( 
\frac{e^2}{\epsilon|\vec r_{ij}|}
-\frac{e^2}{\epsilon \sqrt{(\vec r_{ij})^2+d^2}}
\right)
e^{-\gamma |\vec r_{ij}|} \,.
\end{equation}
Here $\epsilon$ is the dielectric constant
of the substrate, 
and $d/2$ is the distance to the gate plane. 
(The parameter $d$ in (\ref{interaction})
which controls the interaction range 
is chosen in our simulation
in the interval $0 \le d/2 \le 5a$.)
The single dot charging energy $\frac{1}{2}V(0)=e^2/2C$ is assumed to be 
large
enough to inhibit multiple occupancy, so that $q_i = 0,1$. 

The exponential factor in (\ref{interaction}) is introduced for
convenience, to control convergence of the sum in (\ref{E_tot}).
Below we use $\gamma^{-1} = 2d$.
In the case of spatially varying $\epsilon$ the 
form of
interaction can be 
more complicated. 
For example, for an array of dots 
on a 
dielectric
substrate, one 
should replace $\epsilon$ in 
Eq.~(\ref{interaction}) by  $(\epsilon+1)/2$.

Electron tunneling 
in the system \cite{Bawendi,qdots-photo,qdots-transport} presumably 
occurs mainly between neighboring dots.
The tunneling is 
probably incoherent,
i.e. assisted by some energy relaxation mechanism, 
such as phonons. 
Since the tunneling coupling of the dots 
is weak, \cite{Bawendi,qdots-photo,qdots-transport}
we focus on 
the
charge states and ignore 
the
effects of electron spin, such as exchange, spin ordering, etc.

The incoherent nature of electron hopping warrants employing 
stochastic MC
dynamics in which transitions between 
different states take place in accordance with Boltzmann probabilities.
The states undergoing the MC dynamics are 
charge configurations with $q_i=0,1$ 
on a $N\times N$ patch of a triangular array.
Periodic boundary conditions are imposed by 
extending the $N\times N$ 
configurations $q_i$
along with the Hamiltonian (\ref{E_tot})
periodically 
over the entire plane.
Spatial periodicity
of the MC dynamics is 
maintained by
allowing  charge hops across
the boundary, so that the charges disappearing on one side of the patch 
reappear on the opposite side.

Charge conservation gives 
a constraint $\sum q_i = {\rm const}$
that has to be enforced throughout the MC evolution.
While contributing to MC dynamics slowing down,
this constraint 
has no effect on the statistical equilibrium and the 
thermodynamic properties.
To study ordering one can
employ the nonconserving A dynamics, with the 
the gate voltage $V_{\rm g}$ serving as a
parameter controlling 
the system state. This is beneficial due to relatively high speed of the 
A dynamics.
In turn, the somewhat slower charge conserving
B dynamics is used to investigate conductivity at fixed 
charge density
\be \label{charge-density}
n = A^{-1} \sum_i q_i \,, \quad A = N^2 \,. 
\ee
Below we recall the definition of the dynamics A and B, and introduce 
the charge-spin mapping which will be used throughout the paper.

\subsection{The dynamics of type A}

In the MC dynamics of type A, since charge is not conserved, the charge state
is updated independently on different sites. 
The occupancy of 
a randomly selected site $i$ is changed or preserved 
with the probabilities $W_{i}$ and $\overline{W}_{i}$,
which depend on the system state and its attempted change as follows:
\bea 
\label{probabilitiesA}
W_{i}/\overline{W}_{i} & = & \exp(-\delta E_i^{(A)}/T_{\rm el}) \,, \\
\delta E_i^{(A)} & = & \delta q_i \Phi_i  \,,
\label{EA}
\eea
where $W_{i}+\overline{W}_{i}=1$. 
Here $\delta q_i$ is the attempted change of the charge at the site $i$,
whereby the potential is
\be \label{eq:Phi_i}
\Phi_i=\sum\limits_{r_j\ne r_i}
V(\vec r_{ij})q_j + V_{\rm g} + \phi(\vec r_i) \,,
\ee
and $T_{\rm el}$ is the temperature in the charge system
(for brevity, from now on we set $k_B \equiv 1$).
At each MC time step, the charge configuration
is affected only on one site. 
The potential $\Phi_i$ at this site
is obtained using the current system state modified 
at the preceding MC step.

\subsection{The dynamics of type B}

To enforce charge conservation appropriate for
the B dynamics, we update the state of the system
by exchanging charges on randomly chosen neighboring sites.
Specifically, we randomly select 
a site $i$. Then, at each MC time step, we randomly chose
a site $j$ neighboring to $i$ until the occupancy of the 
sites $i$ and $j$ are not the same,  $q_{j} \neq q_i$.
After that, the occupancies of the sites $i$ and $j$ 
are exchanged with probability $W_{i\to j}$, 
and remain unchanged with probability  $W_{i\to i}$, such that
\bea
\label{probabilitiesB}
&& W_{i\to j}/W_{i\to i} = \exp(-\delta E_{ij}^{(B)}/T_{\rm el}) \,, 
\\
\delta E_{ij}^{(B)} &=& (q_j - q_i) (\Phi_i - \Phi_j) - (q_j - q_i)^2 V(\r_{ij}) \,,
\label{EB}
\eea
where $W_{i\to j}+W_{i\to i}=1$,
and the potential $\Phi_i$ is defined by Eq.~(\ref{eq:Phi_i}).

\subsection{Charge-spin mapping}
\label{sec:charge-spin}

The model 
(\ref{E_tot}),(\ref{interaction}),(\ref{probabilitiesA}),(\ref{probabilitiesB}) 
possesses an electron-hole symmetry,
exhibited by introducing a `spin' variable 
\be \label{spin-var}
s_i = 2q_i -1  = \pm 1 \,. 
\ee
The charge Hamiltonian (\ref{E_tot}), rewritten in terms of the spin variables
(\ref{spin-var}), becomes
\bea 
\label{H-s}
%\matrix{ 
{\cal H}_{\rm el} &=& \frac14 {\cal H}_{\rm s}  + {\rm const} \,, \\
{\cal H}_{\rm s} &=& \frac{1}{2}\sum\limits_{i,j}V(\vec r_{ij})s_i s_j
+\sum\limits_{\vec r_i} (\mu + 2\phi(\vec r_i))  s_i \,.
\label{def-Hs}
\eea
Here we introduced the chemical potential  
\be \label{mu}
\mu = 2 V_{\rm g} + V_{{\bf k} = 0} 
\ee
that corresponds to an external field for the spins $s_i$.
Here  $V_{{\bf k}}=\sum_j e^{i\vec r_j\vec k}V(\vec r_j)$ is
the Fourier transform of the interaction (\ref{interaction}).
In terms of spin variables the charge density (\ref{charge-density})
is given by
\be \label{n-s}
n =  A^{-1} \sum_i {1 \over 2}(s_i + 1) \,, \quad A = N^2 \,.
\ee
In the corresponding stochastic dynamics for the spin system, 
defined as above, we use $s_i$ instead of $q_i$ in the energies 
(\ref{EA}) and (\ref{EB}), 
and rescale $T_{\rm el}$ to the spin temperature
\be \label{def-T}
T = 4\, T_{\rm el} \,.
\ee

Eqs.~(\ref{spin-var}) - (\ref{def-T}) 
provide a mapping of the charge problem onto 
a spin system with long-range interaction (\ref{interaction}).
The positive sign of the coupling (\ref{interaction}) 
corresponds to 
antiferromagnetic nearest neighbor interaction.
We note that the limit 
$d \ll a$
corresponds to the \tiafm problem\cite{IsingAFM,Blote82,Blote84} 
with nearest neighbor interaction.
Unless explicitly stated, below we consider
a system without disorder, $\phi(\vec r)=0$.

%%%%%%%%%%%%%%%%%%%%%%%%%%%%%%%%%%%%%%%%%%%%%%%%%%%%%%%%%%%%%%%%%%%%%%
\section{Phase diagram}
\label{sec:phase-diagram}

\subsection{Cooling curves}

\nin
To explore the ground states as a function of the chemical potential $\mu$,
we use the type A (nonconserving) MC dynamics.
To reach the true equilibrium at low temperatures the usual precautions 
are taken by running MC first at an elevated temperature, 
and then gradually decreasing it to the desired value. 

\begin{figure}[t]
\includegraphics[width=3.5in,height=3in]{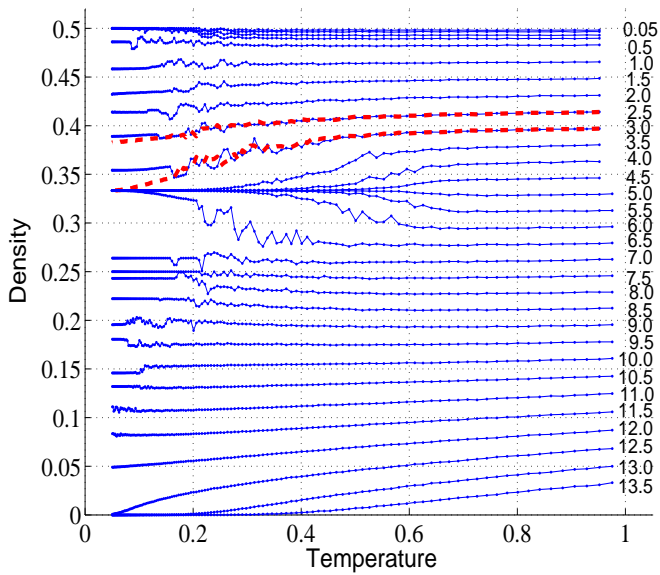} %densd2.eps
\caption[]{(Color online) 
Cooling curves $n(T)$ at fixed gate voltage $V_{\rm g}$
for the screening length parameter $d=2a$. 
Due to the electron-hole symmetry, only the densities $0\le n \le 1/2$ are shown.
The values of $\mu$, related to $V_{\rm g}$ via (\ref{mu}), 
are given on the right side of the plot. Not labelled are the curves with
$\mu = 0.1, 0.2, 0.3$ converging to $n=1/2$.
Additional cooldowns are shown for $\mu=2.5$ and $\mu=3.0$ (dashed line).
    }
 \label{cooling-curves-d2}
\end{figure}

\begin{figure}[t]
\includegraphics[width=3.5in]{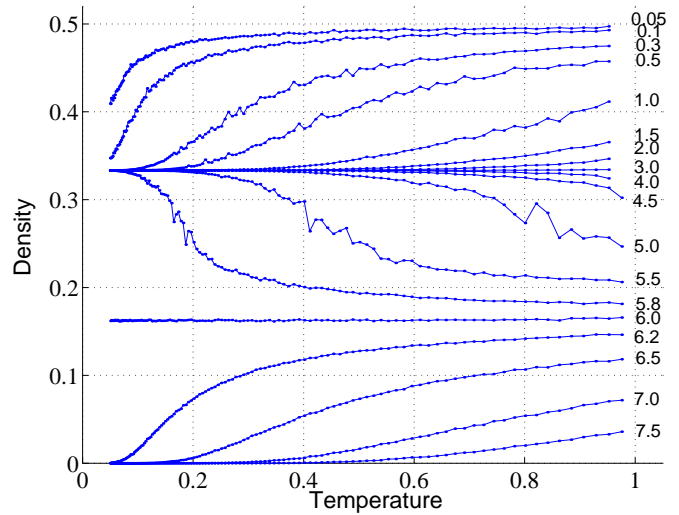} %densising.eps
 \caption[]{(Color online) 
A family of cooling curves 
obtained in the same way as in Fig.~\ref{cooling-curves-d2} 
for $d \ll a$ (the pure \tiafm case). 
The values of $\mu$ are given on the right.
    }
 \label{cooling-curves-ising}
\end{figure}

The cooling curves in Fig.~\ref{cooling-curves-d2}
show the temperature dependence of electron density $n(T)$ for 
a moderate value of the screening length $d=2a$,
with the temperature $T$ measured in units of the nearest neighbor
coupling $V(a)$.
Due to the electron-hole
symmetry $n \leftrightarrow 1-n$, it suffices to consider only  
$0\le n \le 1/2$. 
Similar curves obtained 
for the \tiafm problem realized at small $d \ll a$ are displayed in
Fig.~\ref{cooling-curves-ising}.

%%%%%%%%%%%%%%%%%%%%%%%%%%%%%%%%%%%%%%%%%%%%%%%%%%%%%%%%%%%%%%%%%%%%
%%%%%%%%%%%%%%%%%%%%%%%%%%%%%%%%%%%%%%%%%%%%%%%%%%%%%%%%%%%%%%%%%%%%

The two families of cooling
curves are qualitatively similar in the character of 
temperature dependence: slow at high $T$, faster at lower $T$,
and exhibiting strong fluctuations
before final stabilization at the $T=0$ value. Also, in both cases 
the trajectories
are attracted to the densities $n=0,\,1$ and $n=1/3,\,2/3$.
We note, however, qualitatively new features in the 
case of long range interaction (Fig.~\ref{cooling-curves-d2}).
The most obvious one is that the 
values of $n$ attained at $T\to 0$ span a range of intermediate
densities,
with $n$ being a continuous function of $V_{\rm g}$.
This behavior, indicating freezing transitions at all density values, 
should be contrasted with the $T\to 0$ behavior in the \tiafm case, 
characterized by discontinuous jumps between 
$n = 0$, $1/3$, $2/3$, $1$. 
The latter four values correspond to the
incompressible states [plateaus in the dependence
of $n$ \emph{vs.} $V_{\rm g}$].

%%%%%%%%%%%%%%%%%%%%%%%%%%%%%%%%%%%%%%%%%%%%%%%%%%%%%%%%%%%%%%%%%%%%
%%%%%%%%%%%%%%%%%%%%%%%%%%%%%%%%%%%%%%%%%%%%%%%%%%%%%%%%%%%%%%%%%%%%

The long range interactions
give rise to
ordering at the densities not realized in the \tiafm problem, 
the simple fraction
$n=1/2$
being the most prominent one 
(Fig.~\ref{cooling-curves-d2}).
This density is an attractor for a family of cooling curves at
$|\mu|\lesssim 0.5$: Several such curves, 
obtained for $\mu = 0.05$, $0.1$, $0.2$, $0.3$,
are shown in the top part of Fig.~\ref{cooling-curves-d2}.
Evidently, the basin of attraction 
of the $n=1/2$ state is 
considerably smaller than that of the $n=1/3$ and $n=2/3$ states,
which is to be expected, since the ordering at $n=1/2$ 
is controlled by the next-to-nearest neighbor interactions.

The result of cooling, in general, is found to depend somewhat
on the cooling history, especially near the incompressible densities. 
For different initial random distributions of charge,
and depending on the specific sequence of MC moves,
MC relaxation can lead to 
different ground states. This happens because for a generic long-range
interaction there are many states nearly degenerate in energy,
as illustrated in Fig.~\ref{cooling-curves-d2} by the two 
pairs of curves starting at $\mu=2.5$ and $\mu=3.0$ 
obtained for different runs of the MC dynamics.
These curves, identical at high $T$, 
diverge below the temperature interval where fluctuations develop.
The dependence on the cooling history 
in the fluctuation region limits the accuracy of phase diagram obtained
from cooling trajectories in a finite system.

\begin{figure}[b]
\includegraphics[width=3.5in]{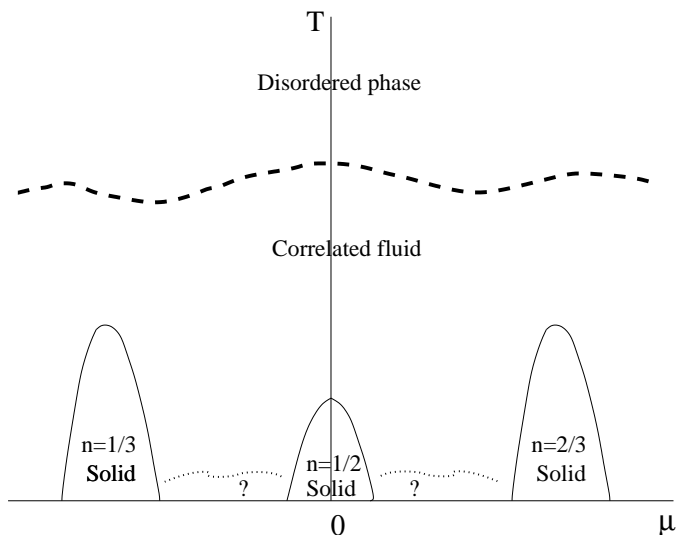} %phasediagram
 \caption[]{ 
Schematic phase diagram on the $(\mu, T)$ plane for 
a generic value of the screening length parameter $d > 0$.
The dashed line marks the crossover between the correlated fluid and the disordered
phase. 
The low temperature phases denoted by a question mark could be either commensurate
or disordered. 
}
 \label{phasediagram}
\end{figure}

Figure~\ref{phasediagram} 
summarizes in a schematic way the results of the MC study of cooling.
It shows the phase diagram of the system in the  
$(\mu, T)$ plane. Due to the particle-hole equivalence,
it posesses the $\mu \leftrightarrow -\mu$ symmetry. 
For a generic value of the chemical potential we find
three distinct temperature phases: 
the {\it disordered} state at high temperature, 
the {\it correlated fluid} phase at intermediate temperatures, and
the {\it solid} 
phases at low $T$.

%%%%%%%%%%%%%%%%%%%%%%%%%%%%%%%%%%%%%%%%%%%%%%%%%%%%%%%%%%%%%%%%%%%%
%%%%%%%%%%%%%%%%%%%%%%%%%%%%%%%%%%%%%%%%%%%%%%%%%%%%%%%%%%%%%%%%%%%%

\subsection{Freezing transitions}
\label{sec:freezing}

\nin 
Ordering at different $n$ resembles, and indeed can be connected to, 
the phase transitions in adsorbed monolayers.\cite{denNijs}
The latter have been mapped on the known statistical models
(Ising, Potts, etc), some of which are exactly solvable.\cite{Baxter-book} 
Possible phase transitions in 2d 
have been classified by Domany {\it et al.}\cite{Domany}
and by Rottman\cite{Rottman} based on the Landau theory.\cite{Landau5}

The new ingredient in the charge system, the long range interaction,
does not change the symmetry of the 
Landau free energy.\cite{Domany,Rottman}
Still, the long range interaction can, in principle, 
change the order of the transition if the latter is dominated 
by fluctuations, making it deviate from the Landau theory scenario.
While determining the specifics of the freezing transitions 
in this more general
case is beyond the scope of the present work, 
one can make the following observations.

(i) At the densities $n=1/3$, $2/3$, freezing occurs into one of the three 
degenerate $\sqrt{3}\times\sqrt{3}$ configurations 
[see Fig.~\ref{snapshots} (A) below]. In this state, electrons can
occupy one out of the three sublattices of the triangular lattice.
We believe that this transition is of the first order.
Our argument is based on the step-like singularity of the average energy of 
the system at the transition observed during the MC dynamics,\cite{unpub}
and is corroborated by a fairly sharp step-like singularity in the MC conductivity
(see below, Section~\ref{sec:conductivity}), interpreted using the general
connection of the singularities in conductivity 
and in average energy.\cite{Hohenberg-Halperin}
Besides, for this transition a cubic invariant in the Landau
free energy is allowed.\cite{Domany,Rottman,unpub}

An alternative scenario for freezing at this density is a continuous transition
of the $q=3$ Potts universality class.\cite{Domany,Rottman}
This possibility, in principle, cannot be ruled out based just on symmetry, since 
in two dimensions there are exceptions from the Landau theory, 
notably the $q=3,4$ Potts models, for which the transition
is second-order even though the cubic invariants exist.\cite{Baxter-book}
We believe, however, that in the present case the existence of the cubic 
invariant triggers the first order transition.
%since the long-range interaction (\ref{interaction}) 
%suppresses fluctuations, justifying validity 
%of the Landau theory. 
We also note that the experimental evidence 
for adsorbtion of nitrogen molecules on graphite does not contradict 
the first-order transition scenario\cite{N2-C-exp} 
even for fairly short-ranged quadrupolar molecular interactions.

(ii) At the density $n=1/2$, 
the ground state is a $(2\times1)$ charge density wave
of three possible orientations [see Fig.~\ref{snapshots} (C)].
In the ordered state electrons occupy one out of the two sublattices,
resulting in the six-fold degeneracy of the ground state.

The nature of the phase transition for $n=1/2$ is less clear.
The classification\cite{Domany,Rottman} suggests 
the $q=4$ Potts universality class with a second-order transition.
However, the electron-hole symmetry at $\mu=0$ line forbids the cubic invariant 
in the corresponding Landau free energy, putting this transition into 
the universality class of 
the Heisenberg model with cubic anisotropy.\cite{Domany-Riedel}
The transition in the latter model 
is still poorly understood,\cite{cubic-Heisenberg}
with both the continuous and the fluctuation-induced first order transition
on the table.\cite{Calabrese}
Recent study\cite{Korshunov'05} of the zero-field \tiafm with finite-range 
interactions (beyond the nearest-neighbor) generally
favors the first order transition, 
while leaving room for an additional continuous transition 
at special values of couplings.
At $\mu\ne 0$, in the absence of electron-hole symmetry,
the presence of the cubic invariant makes
the first order transition scenario even more likely.

Our numerical accuracy does not allow us to make a definite prediction.
The observed kink in the conductivity (Section~\ref{sec:conductivity} below)
is consistent with either the first or the second order transition,
as is the singularity in the average energy.\cite{unpub}
In general, freezing transition into the $n=1/2$ ground state is less
pronounced than that for $n=1/3$, $2/3$ states, since it is determined
by the next-to-nearest neighbor coupling.

(iii) For a generic density $n$ we observe that 
at decreasing $T$ the MC dynamics slows down, and 
all the charges eventually become immobile. 
The ground state in general looks disordered, and depends somewhat 
on the cooling
history. The system configuration
space appears to have many nearly degenerate minima, 
which complicates finding the 
true ground state numerically.
While a disordered ground state 
for a continuous range of densities
cannot be ruled out,
we anticipate freezing into commensurate
``epitaxial solids'' for at least some rational 
densities $n=p/q$ with higher denominators, such as the 
striped ground state for the density $n=3/7$ shown in Fig.~\ref{snapshots} (D).
These transitions would take place 
at ever smaller temperatures since they are governed
by the couplings $V(\r)$ beyond nearest and next-to-nearest.
For instance, for our model interaction (\ref{interaction}) with $d=2a$,
freezing into the $n=3/7$ state occurs at $T\sim 0.01V(a)$, and requires
about 10 hours of CPU time.
We comment on possible freezing scenarios 
in Section~\ref{sec:discussion} below.

%%%%%%%%%%%%%%%%%%%%%%%%%%%%%%%%%%%%%%%%%%%%%%%%%%%%%%%%%%%%%%%%%%%%%%%%%%
%%%%%%%%%%%%%%%%%%%%%%%%%%%%%%%%%%%%%%%%%%%%%%%%%%%%%%%%%%%%%%%%%%%%%%%%%%
\section{Charge Transport:  Electrical Conductivity}
\label{sec:conductivity}

\nin
Our interest in the hopping transport is two-fold.
First, the conductivity is experimentally accessible 
in the dot arrays.\cite{qdots-photo,qdots-transport}
Second, as we shall see below, the dc conductivity 
is sensitive to the thermodynamic state, and its temperature and density 
dependence can thus be used to distinguish between different phases 
of the system.

Below we use the MC procedure to compute the hopping conductivity in 
the presence of a small external electric field. 
The MC conductivity $\sigma_{\rm MC}$ and the one measured in a real system
are related as follows.
Electron hopping between neighboring dots is
assisted by some energy relaxation mechanism, such as phonons. 
The latter adds a temperature-dependent prefactor $f(T)$ to the hopping rate,
so that $\sigma_{\rm total} = f(T) \sigma_{\rm MC} (T)$. 
[In a simple model involving coupling to acoustic phonons, 
a Golden Rule calculation gives
a power law $f(T)\propto T^{\alpha}$.]
In this work,
for simplicity, we shall ignore the system-dependent prefactor $f(T)$,
and focus on the MC conductivity $\sigma_{MC}$.

%%%%%%%%%%%%%%%%%%%%%%%%%%%%%%%%%%%%%%%%%%%%%%%%%%%%%%%%%%%%%%%%%%%%
\begin{figure}[t]
\includegraphics[width=3.5in]{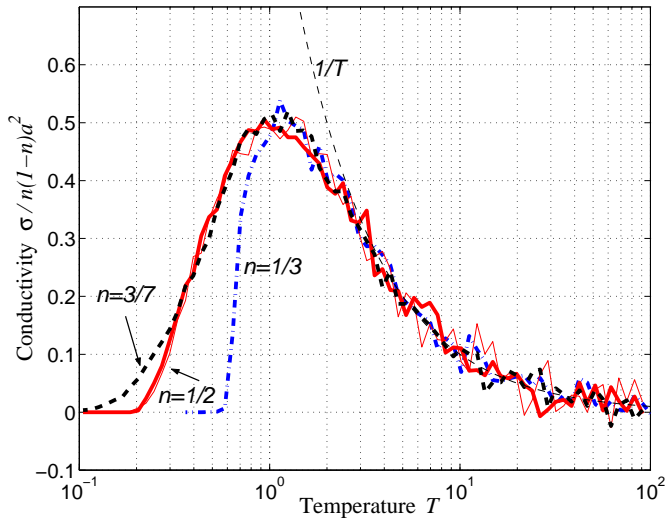} %condvsT.eps
 \caption[]{(Color online) 
Temperature dependence of the zero bias dc conductivity $\sigma(T)$.
Shown are the curves $\sigma/n(1-n)$ for $n=1/3$, $n=1/2$, 
and for a typical intermediate density 
(taken here to be $n=3/7$), obtained for $18\times 18$ system
with the screening length $d=2a$.
The asymptotic large temperature behavior 
(\ref{sigma-high-T}) is indicated by dashed line.
The faint solid line, 
describing $\sigma(T)$ for $n=1/2$ on $12\times 12$ patch,
coincides with that for the $18\times 18$ system, 
indicating the smallness of finite size effects.
    }
 \label{condvsT}
\end{figure}
%%%%%%%%%%%%%%%%%%%%%%%%%%%%%%%%%%%%%%%%%%%%%%%%%%%%%%%%%%%%%%%%%%%%

The temperature dependence of $\sigma_{MC}$ 
is shown in Fig.~\ref{condvsT} for several densities. 
The simulation was performed on a $18\times 18$
patch using the charge-conserving dynamics (type B). 
The external field $\E$, applied along the patch side, 
is chosen to be large enough 
to induce a current measurable in the presence of thermal fluctuations,
and yet sufficiently small to ensure the linear response 
\be \label{def-cond}
\j = \sigma\E \,.
\ee 
The data was obtained using $E$ in the range
\[
E a  \simeq ( 10^{-2} - 5\times 10^{-2}) \, V(a) \,. 
\]
We observed that the linearity holds 
for the field values much smaller
than both the temperature 
and the next-to-nearest neighbor interaction, 
$E a  \ll {\rm min} \{V_{nnn}, T\}$
[Eq.~(\ref{eq:VnnVnnn})].

The conductivity $\sigma$, Eq.~(\ref{def-cond}), 
is obtained using the MC dynamics of type B as follows.
Since the field contribution to the
potential difference between the adjacent sites separated by 
${\bf a}=\r_{ij}$, 
% $|{\bf a}| = a$, 
is $-{\bf Ea}$, 
we can incorporate the effect of the field by adding the quantity
\be \label{deltaE-field}
(q_i - q_j) {\bf Ea}
\ee
to the energy difference $\delta E_{ij}^{(B)}$ in Eq.~(\ref{EB}).

Our simulation was carried out using the spin representation,
as described in Sec.~\ref{sec:charge-spin}. 
In the spin language, using the Hamiltonian
(\ref{def-Hs}) and the temperature (\ref{def-T}), 
the term corresponding to Eq.~(\ref{deltaE-field}) is $(s_i-s_j){\bf Ea}$.
The data presented below was obtained using 
the field $\E$ oriented along the height of 
an elementary lattice triangle, so that $|{\bf Ea}| = Ea\cos \frac{\pi}6$. 
The `spin' current density was 
calculated as 
\be \label{def-j-cond}
j = |\delta s| a\cos(\pi/6) \cdot 
{{\cal N}_+ - {\cal N}_- \over {\cal N}} , 
\ee 
where ${\cal N}_{\pm}$ is the number of hops along
(against) the direction of $\E$
during the MC run time,  
$\cal{N}$ is the total number of MC trials at each temperature step,
and $\delta s=\pm 2$ is the spin change for each hop.
The corresponding charge current is then $j_{\rm el}= \frac12 j$.

The MC conductivity as a function of temperature is displayed 
in Fig.~\ref{condvsT} for several values of charge density. 
From the dependence $\sigma(T)$ one can identify three 
temperature intervals with different behavior, corresponding to
the high, low, and intermediate temperatures. 
Relevant temperature scales
are approximately given by the nearest neighbor
and next-to-nearest neighbor interaction strength:
\be\label{eq:VnnVnnn}
V_{nn}=V(a),\quad V_{nnn}=V(\sqrt{3}a).
\ee
In our MC simulation, for $d=2a$, the value of 
$V_{nnn}$, given by the interaction across the main diagonal of a rhombus, 
was about $0.3V_{nn}$.

The simplest to understand is the high temperature
behavior $T\ge V_{nn}$, corresponding to a
{\it disordered} phase in which conductivity takes place via
uncorrelated hops of individual electrons. 
Conductivity in this phase 
has a simple temperature and density dependence 
[see Eq.~(\ref{sigma-high-T})].

\begin{figure}[b]
\includegraphics[width=3.5in]{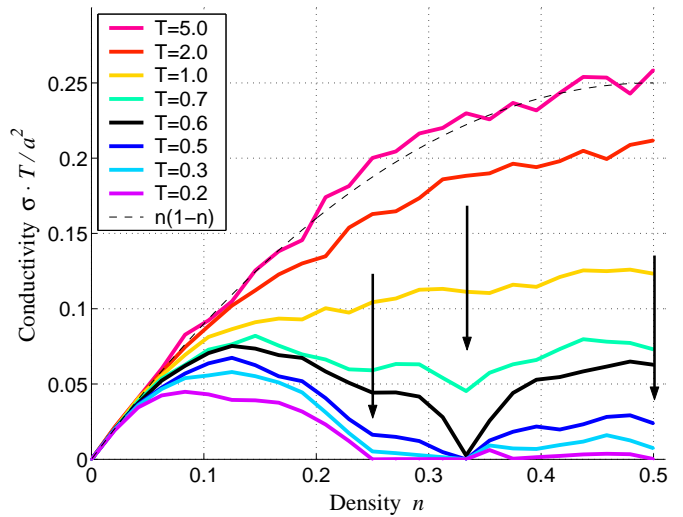} %T-n-plot.eps
%\centerline{\psfig{file=T-n-plot.eps,width=3.5in}}
%\vspace{0.5cm}
 \caption[]{(Color online) 
The conductivity $\sigma$, scaled by $a^2/T$, is shown as a function of
electron density for several temperatures.
The temperature values are given in the units of $V(a)$;
the interaction $V(\r)$ of the form (\ref{interaction}) was used 
with the screening length $d=2a$. 
Arrows mark the features corresponding to the freezing phase transitions 
at $n=1/4$, $1/3$, $1/2$.
Dashed line 
corresponds to
the high temperature limit 
described by
Eq.~(\ref{sigma-high-T}).     
}
\label{T-n-plot}
\end{figure}

In the opposite limit,
at $T\to 0$, the system freezes into the ground state
configuration ({\it solid} phase), and the conductivity
vanishes. The freezing transition is entirely due to the 
longer range coupling, such as $V_{nnn}$,
since for purely nearest neighbor coupling,
realized in the \tiafm model, the system 
does not exhibit a phase transition and
is characterized by finite entropy even at $T=0$.
Thus the upper temperature scale for the solid phase can be estimated
as $T\lesssim V_{nnn}$. 
The ground state
depends on the density $n$ in a complicated way.
Near rational $n$ the ground state is commensurate,
while at a generic $n$ the state is probably incommensurate. 
The freezing temperature $T_c$ is also a function of the density $n$.
As the temperature approaches the freezing temperature $T_c$,
the conductivity vanishes (Fig.~\ref{condvsT}). 
While the behavior $\sigma(T)$ near $T=T_c$ appears to be singular
(see Fig.~\ref{condvsT}), we were not able to extract 
the exact form of this singularity directly from our simulation.
However, some information about the singularity in conductivity 
can be obtained from general arguments \cite{Hohenberg-Halperin} relating
it with the distribution of energies
studied in Sec.~\ref{sec:phase-diagram}.
We also observe
singularities near rational $n$ 
in the conductivity dependence on the density,
$\sigma(n)$. They are indicated by arrows in Fig.~\ref{T-n-plot}.

Finally, at densities $1/3 \le n \le 2/3$ 
there is an interesting intermediate temperature interval 
\be\label{eq:Vnn<T<Vnnn}
V_{nnn}\lesssim T\lesssim V_{nn}
\ee
in which the 
conductivity is finite and 
reaches maximum as a function of temperature (Fig.~\ref{condvsT}). 
Transport at these temperatures
is of a collective
character ({\it correlated fluid} phase)
owing to strong short-range correlations between 
charges 
which constrain individual 
charge movements. Due to the frustration,
the onset of short-range ordering 
does not lead to charges freezing. The system appears to possess 
a sufficient amount of residual entropy to allow for finite conductivity.

An independent consistency check for
the MC dynamics 
is provided by 
the fluctuation-dissipation theorem,\cite{Landau5}
relating current fluctuations to conductivity:
\be \label{FDT}
\int \! dt \left< j_{\mu}(t) j_{\nu}(0) \right> = 2 \sigma_{\mu \nu} T \,.  
\ee
Our simulations is found to be in
accord with (\ref{FDT}) in all temperature regions,
which ensures that
the MC conductivity indeed describes transport in the linear response regime.
(Some deviations from (\ref{FDT}) were observed 
very close to the freezing point, where $\sigma$ becomes very small.)
In the correlated
phase we explicitly evaluate the integral in the left hand side of 
(\ref{FDT}) by numerically averaging  
the product $j_{\mu}(t+\tau)j_{\nu}(t)$
over $\tau$ and $t$.
The result is compared with the conductivity obtained directly from
Eqs.~(\ref{def-cond}),\,(\ref{def-j-cond})
to  make sure that during an MC run the system 
has enough time to reach equilibrium. 

At large temperature $T \gg V(a)$, one can evaluate the 
left hand side of 
the fluctuation-dissipation relation
(\ref{FDT}) explicitly and find the universal
high temperature asymptotic behavior of the conductivity,
\be \label{sigma-high-T}
\sigma  = a^2\; {n(1-n)\over T} \,.
\ee
To obtain (\ref{sigma-high-T}) we note that 
for a high enough temperature the current is delta-correlated in time, 
\be \label{def-j2}
\left< j_{\mu}(t) j_{\nu}(0) \right> = 
\left< j^2 \right> \, \delta_{\mu \nu}\, \delta(t) \,.
\ee
The mean square $\left< j^2 \right>$ can be obtained by summing
the probabilities of possible MC moves:
\be \label{j2-yy}
\left< j^2 \right> = 
{4\over 6}\cdot 2n(1-n)\cdot (\delta s)^2 \cdot \lp a\cos{\pi\over 6}\rp^2
\cdot w
\ee 
The factor ${4\over 6}$ comes about because
in the field $\E$ 
aligned along the height of the elementary lattice triangle, 
only four out of possible six bond directions contribute to conductivity. 
The second factor, $2n(1-n)$, describes the probability to select 
two adjacent sites occupied by an electron and a hole.
The change of occupancy per MC move is $\delta s=\pm 2$, since $s_i=\pm1$.
The expression (\ref{j2-yy})
does not depend on temperature since 
all the hops are equally probable
and uncorrelated,
$W_{i\to j} = W_{i\to i} \equiv w = 1/2$ at $T \gg V(a)$.
The conductivity tensor is isotropic, 
$\sigma_{\mu\nu} = \sigma \, \delta_{\mu \nu}$,
as can be explicitly checked
by doing a similar high temperature calculation 
for an arbitrary orientation of the field $\E$.
Fig.~\ref{condvsT} shows that the results of our MC dynamics
are consistent with the 
high temperature behavior 
(\ref{sigma-high-T}).

%%%%%%%%%%%%%%%%%%%%%%%%%%%%%%%%%%%%%%%%%%%%%%%%%%%%%%%%%%%%%%%%%%%%
%%%%%%%%%%%%%%%%%%%%%%%%%%%%%%%%%%%%%%%%%%%%%%%%%%%%%%%%%%%%%%%%%%%%

%%%%%%%%%%%%%%%%%%%%%%%%%%%%%%%%%%%%%%%%%%%%%%%%%%%%%%%%%%%%%%%%%%%%
\begin{figure}[t]
\centerline{
{\bf A}\includegraphics[width=1.5in]{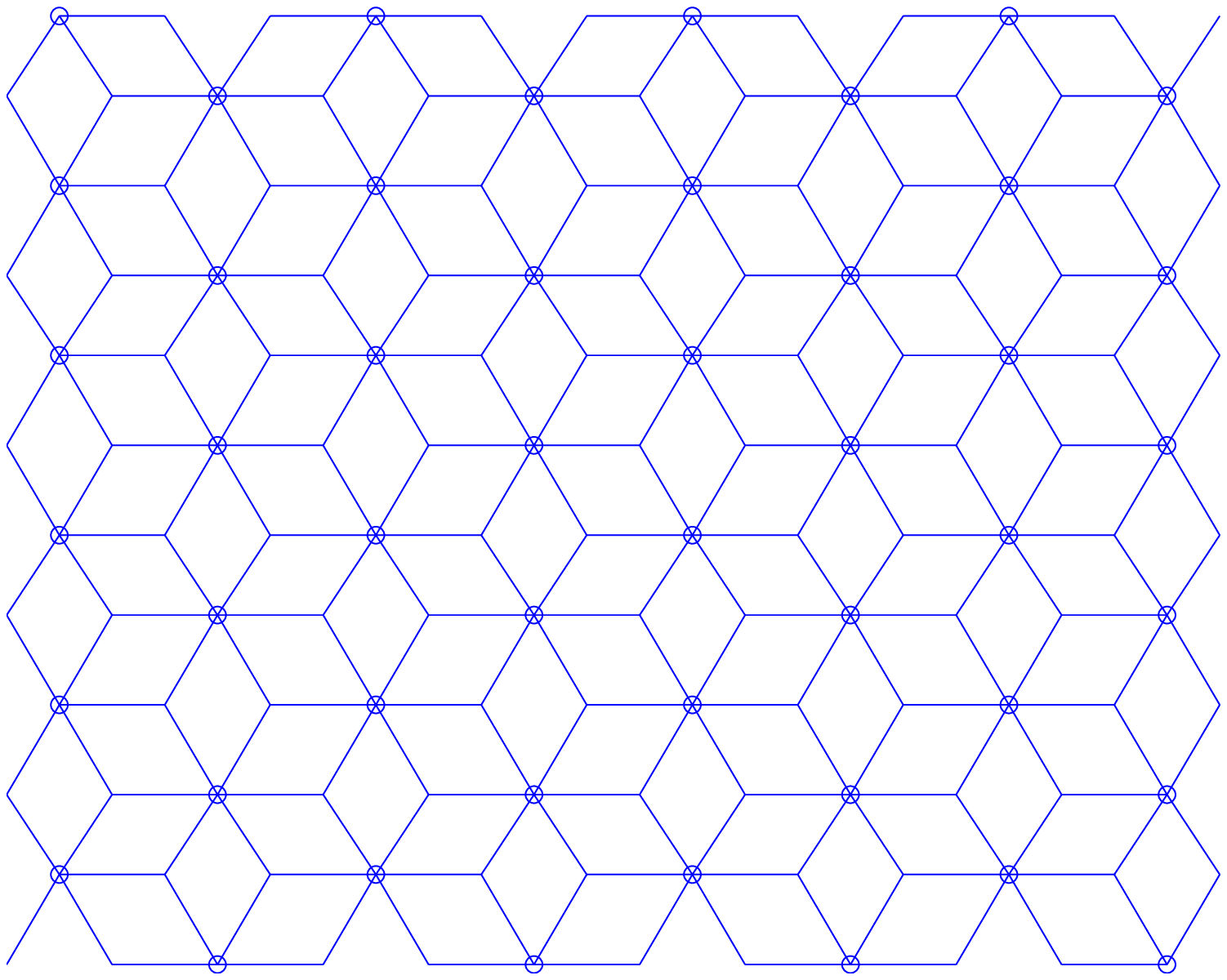} %n48.eps
{\bf B}\includegraphics[width=1.5in]{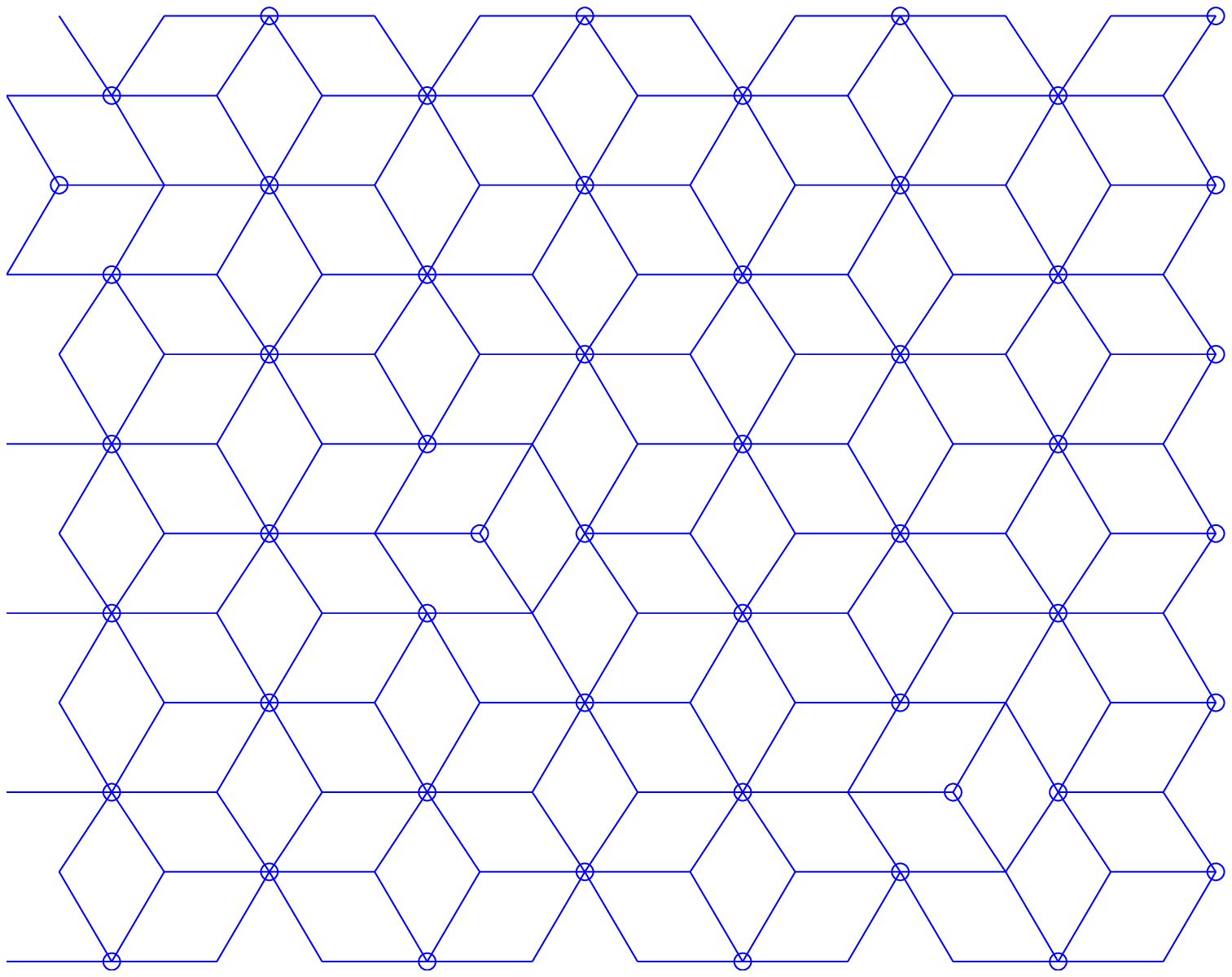} %n51.eps
}
\centerline{
{\bf C}\includegraphics[width=1.5in]{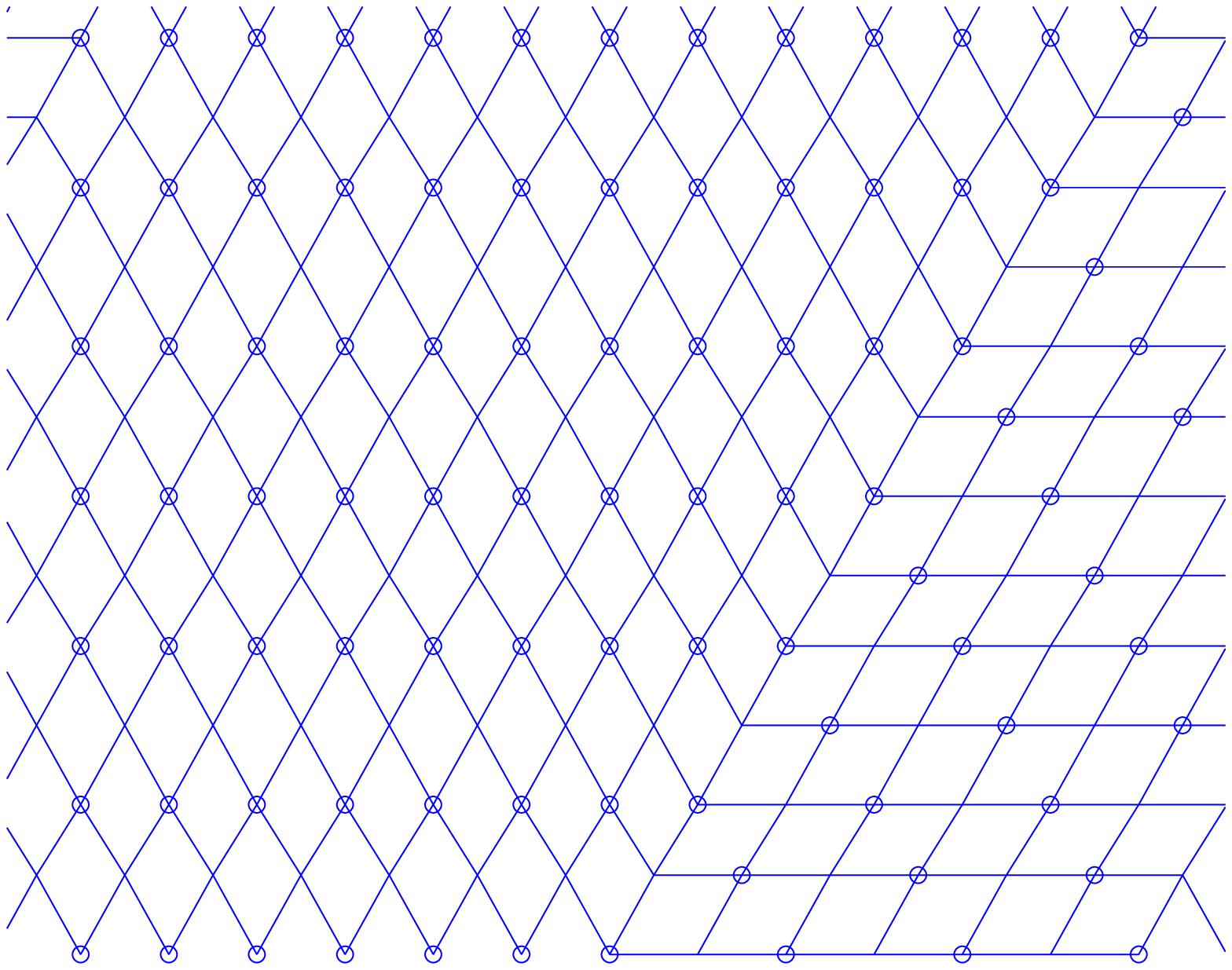} %n12gs.eps
{\bf D}\includegraphics[width=1.5in]{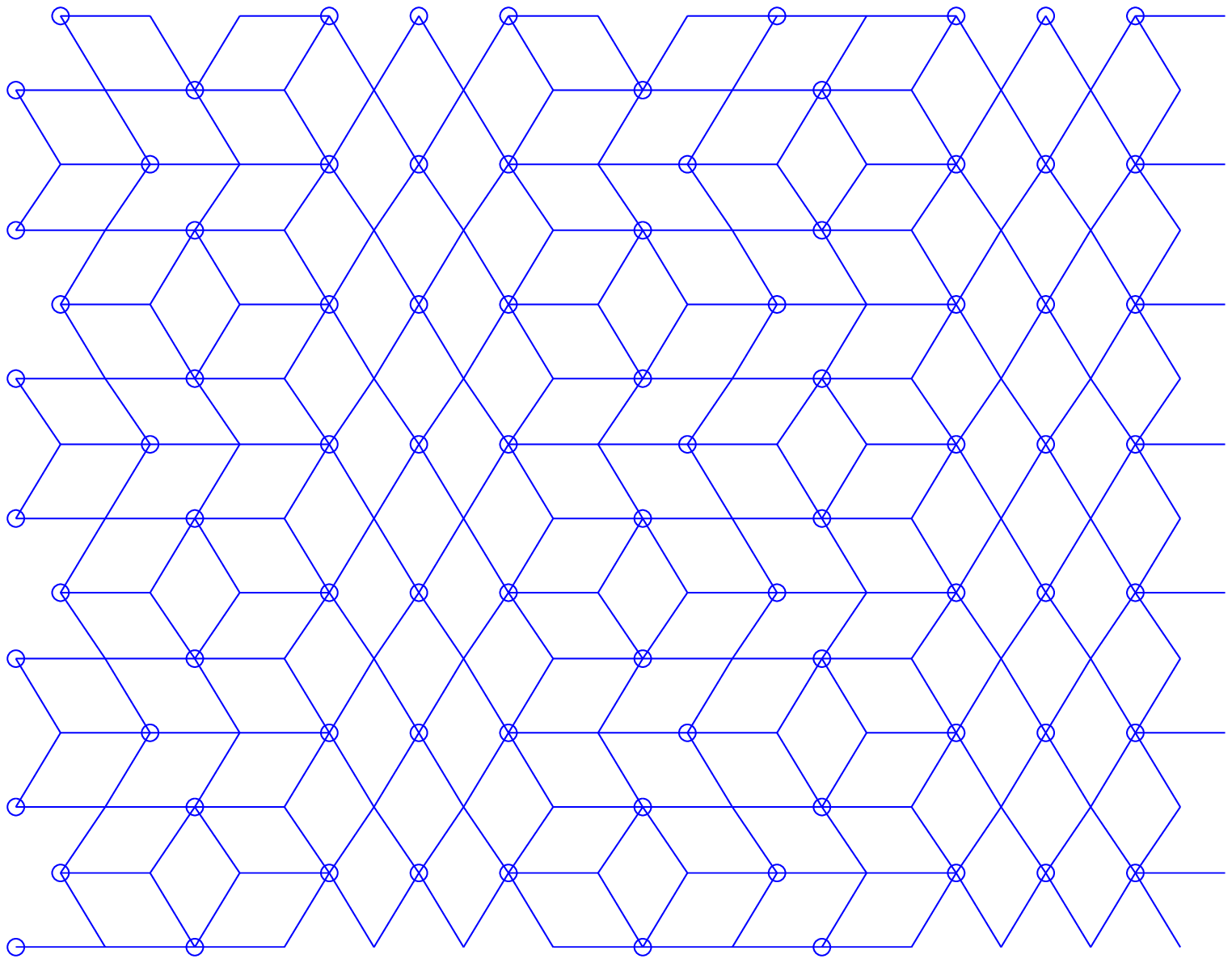} %n42d2gs.eps
}
\centerline{
{\bf E}\includegraphics[width=1.5in]{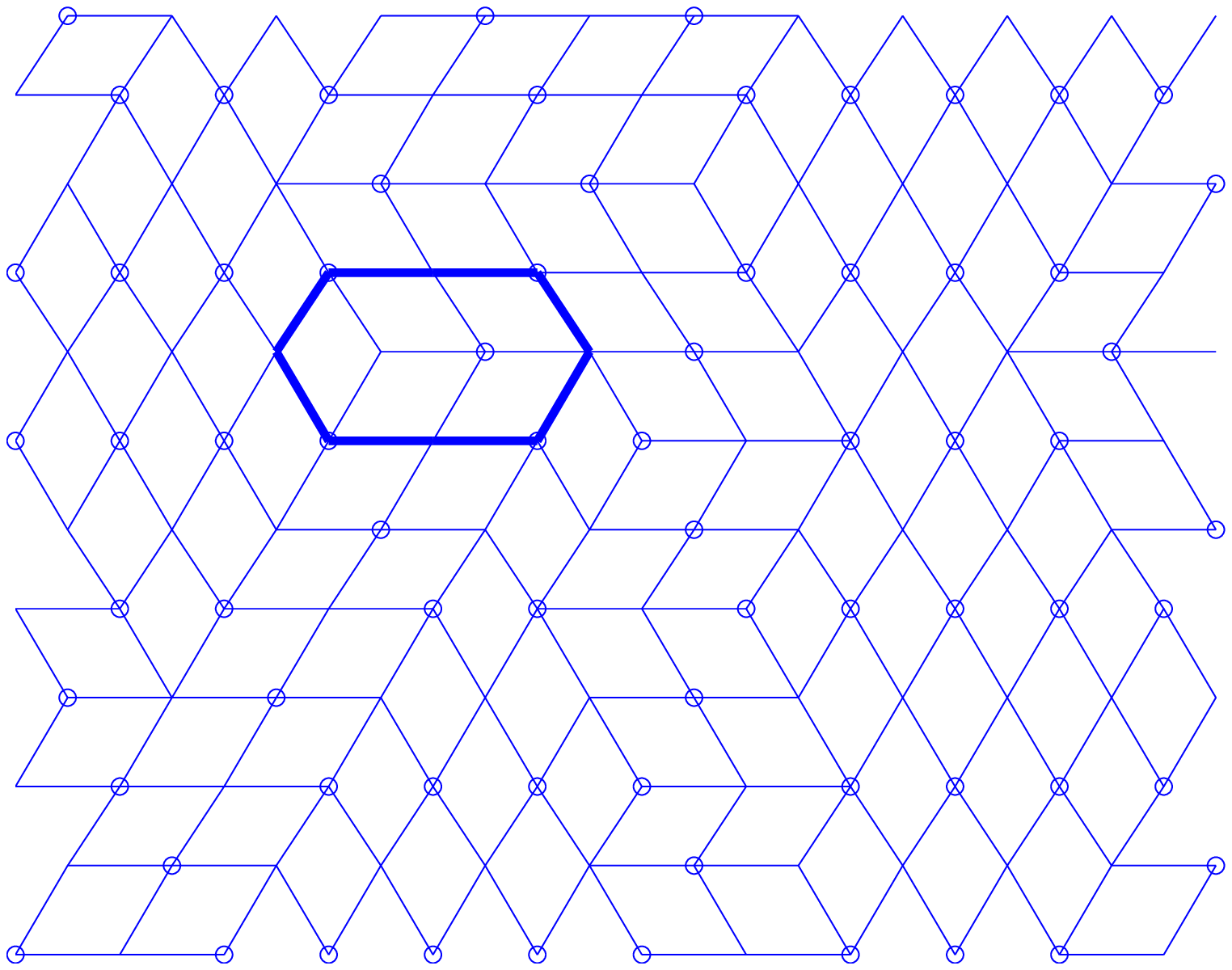} %n72.eps
{\bf F}\includegraphics[width=1.5in]{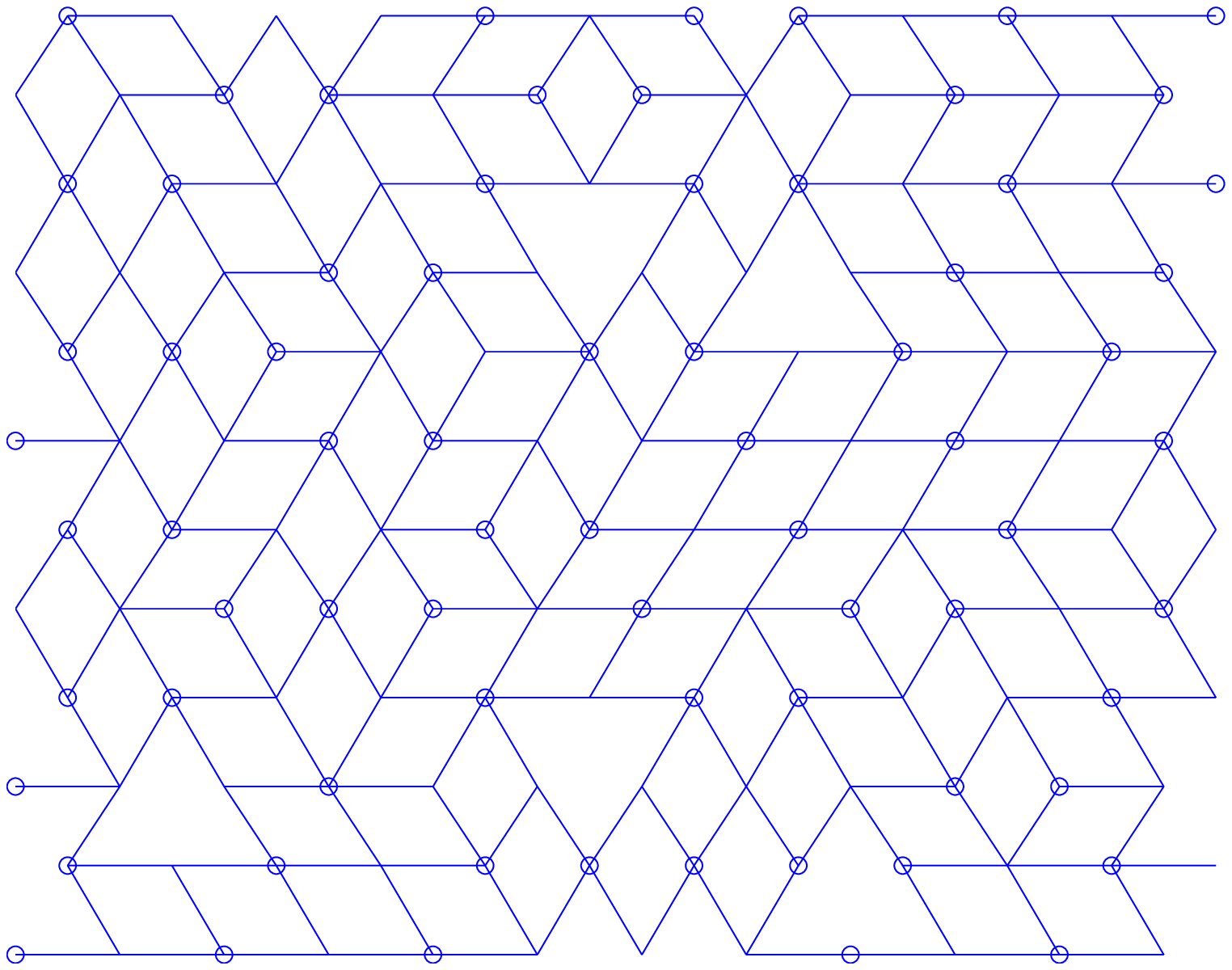} %n72dis.eps
}
%\vspace{0.5cm}
 \caption[]{(Color online) 
        {\bf A:}
Commensurate $\sqrt{3}\times\sqrt{3}$ charge ground state at the density $n=1/3$. 
Electrons are are represented by small dots
on the triangular array sites. Pairing of the triangles is revealed 
by erasing all frustrated bonds connecting the 
sites with equal occupancy, as described in Section~\ref{sec:height-variable}.
        {\bf B:}
Typical charge configuration for $n=1/3+\epsilon$. 
There are three excess charges in the system ($\epsilon=3/144$)
hopping over the honeycomb network of unoccupied sites in the  
state shown in panel A. The excess carriers,
dilute at $\epsilon\ll1$, are moving nearly independently 
on the frozen $\sqrt{3}\times\sqrt{3}$ state background.
        {\bf C:} 
Ground state for $n=1/2$. Shown are the two charge density wave 
domains characterized by different slope orientations [Eq.~(\ref{def-slope})].
        {\bf D:} 
Example of a ground state for a 
commensurate density with a higher denominator, here $n=3/7$.
        {\bf E:} 
Typical charge configuration obtained in a simulation 
for $n=1/2$ in the correlated fluid phase.
Two elementary `3d cubic' cells corresponding to a free charge are 
marked. 
        {\bf F:}
Typical charge configuration for $n=1/2$
at temperature somewhat higher than in panel C. 
In this case, there are several topological defects 
(unpaired triangles) present in the system. 
These defects can be interpreted as dislocations of the height field
(see Sec.~\ref{sec:height-variable}).
}
 \label{snapshots}
\end{figure}
%%%%%%%%%%%%%%%%%%%%%%%%%%%%%%%%%%%%%%%%%%%%%%%%%%%%%%%%%%%%%%%%%%%%

The change in conductivity 
behavior 
while cooling down from the disordered
into the correlated fluid and solid phases 
can also be seen in Fig.~\ref{T-n-plot}.
Here we plot  
the conductivity $\sigma$, scaled by  $a^2/T$,
as a function of the electron density $n$ for
several temperatures. 
The results, shown for the densities
$0\le n\le 1/2$, 
can be extended to $n>1/2$ 
using the electron-hole symmetry $n \leftrightarrow 1-n$.
The high temperature curve is clearly consistent
with the $n(1-n)$ dependence (\ref{sigma-high-T}). 
The low temperature curves 
indicate that the conductivity vanishes 
more quickly
near the densities 
of a simple fraction form ($n=1/4$, $1/3$, $1/2$).
This 
corresponds to
freezing of the system into a commensurate state
at these values of $n$. The commensurate states at $n=1/3$ and $n=1/2$ are shown 
in Fig.~\ref{snapshots} (panels A and C respectively).

At densities near these simple 
fractional values,
the system conducts via hops of excess electrons or holes
moving in the frozen crystalline background. 
Such a situation is 
depicted in Fig.~\ref{snapshots} (panel B)  
for the case of $n=1/3+\epsilon$.
Since the conductivity is proportional to the excess 
charge density $\epsilon$, 
we expect
$\sigma(n)$ 
to have cusps near simple fractional densities. Such cusps are indeed seen
in Fig.~\ref{T-n-plot} near $n=1/4$, $1/3$, $1/2$.

%%%%%%%%%%%%%%%%%%%%%%%%%%%%%%%%%%%%%%%%%%%%%%%%%%%%%%%%%%%%%%%%%%%%%%%%%%
\section{The height variable}
\label{sec:height-variable}

\subsection{The ground state}
\label{sec:ground-state}

\nin
Here we attempt to understand the
ordering in the ground state 
and in the correlated fluid by employing 
the notion of the height field order parameter, originally introduced 
in the context of the \tiafm problem
by Bl\"ote \emph{et al}.\cite{Blote82,Blote84}
In our model (\ref{E_tot}), (\ref{interaction}), 
the latter problem corresponds to the limit $d\ll a$
dominated by the nearest neighbor interactions.
The essential features of the ground state 
can be understood by considering
energy minimization for individual plaquettes (triangles).
With only two kinds of charges it is impossible to avoid the frustrated 
bonds between the like charges, with at least one such bond per triangle.
In the ground state, the number of such bonds must be as small as possible.
That can be achieved by \emph{pairing} the
neighboring triangles in such a way that each pair shares a 
frustrated bond. One can see that the charge state corresponding 
to all triangles paired, while providing the absolute energy minimum, is not unique.
On the contrary, the pairing condition leaves plenty of freedom in the 
charge configuration, characterized by extensive entropy 
(finite entropy-to-area ratio).

The configurations permissible by the pairing condition have 
a simple geometric interpetation. 
After erasing all frustrated bonds, one obtains a rhombic tiling 
of the triangular lattice,\cite{Blote82,Blote84}
as shown in Fig.~\ref{snapshots} (panels A - E). 
It is convenient to view such a tiling as a 2d surface in a 3d cubic crystal
projected along the (111) axis on a perpendicular plane.  
After undoing the projection by lifting the 2d configuration of rhombi
in the 3d space, each site $\vec r_i$ acquires a scalar variable
$h(\vec r_i)$ which is the height of the lifted
surface in a 3d cubic crystal. The field $h$
takes values which are multiples of the distance between
the cubic crystal planes,
\be \label{def-b}
b = {\ell\over 3} = {a \over \sqrt{2}} \, ,
\ee
where $\ell$ is the main diagonal of the unit cell in the cubic crystal.
The mapping onto a continuous height surface
exists only for the densities $1/3 \le n \le 2/3$, 
which we focus on hereafter, since for 
$n$ outside this interval the tiling contains voids.

As we discussed 
above, the long range character 
of the interaction (\ref{interaction}) lifts the ground state degeneracy,
leading to freezing into specific ground states.
To use the height variable 
to characterize the ordering,
we define the {\it slope} ${\bf t}$ of the height surface. 
For that we consider the surface normal vector
\be \label{def-m}
{\bf m} = {\A^{-1}}\, \sum_{s=1}^{3} \, r_s {\bf e}_s \,, \quad
\A = r_1+r_2+r_3  
\ee
where $r_1$, $r_2$, $r_3$ are the numbers of rhombi formed by 
$({\bf e}_2,{\bf e}_3)$, 
$({\bf e}_3,{\bf e}_1)$, $({\bf e}_1,{\bf e}_2)$, 
which add up to the total area ${\cal A}=N^2$.
The slope ${\bf t}$ is given by the projection
\be \label{def-slope}
{\bf t} = {\bf m} - \lp \hat{\bf z}\cdot {\bf m}\rp \hat{\bf z} \,, \quad 
\hat{\bf z} = {1\over \sqrt{3}} \,
\lp {\bf e}_1 + {\bf e}_2 + {\bf e}_3 \rp \,.
\ee
Here the vectors ${\bf e}_s$, $s=1,2,3$ form the basis of the auxiliary 3d crystal
(Fig.~\ref{fig:coordinates} in the Appendix), and 
$r_s$ give the numbers of 3d crystal faces normal to the vectors ${\bf e}_s$.
Eq.~(\ref{def-slope}) defines an {\it average slope} of the 
configuration.
For a `smooth' surface, the vector ${\bf t}(\r)$ is proportional to 
the local height gradient, ${\bf t}(\r) \propto (\partial_x h, \partial_y h)$.

When the interactions are of the nearest neighbor kind ($d\ll a$), 
the height surface fluctuates freely even at small $T$.
We observe a similar behavior at {\it finite} temperatures (above freezing)
for the system with the long range interaction (\ref{interaction}).
The numbers $r_s$, $s=1,2,3$, for an $N\times N$ patch 
are equal on average,
fluctuating 
around the mean value of $N^2/3$, and thus
the average slope is ${\bf t}={\bf 0}$.
As $T\to 0$, for the studied densities $n=1/3$, $2/3$, and $n=1/2$,
the interaction (\ref{interaction}) leads to freezing 
into commensurate ground states characterized by specific discrete slopes
and a non-extensive entropy.

\subsection{The correlated fluid}
\label{sec:corr-fluid}

\nin
The height field $h$ is defined globally and uniquely 
(modulo an additive constant and an overall sign) 
for any of the degenerate ground states of the \tiafm model.
For $T>0$, however, due to thermal fluctuations,
some of the triangles are left unpaired. Such unpaired triangles 
represent topological defects of the height field. 
This can be seen most clearly at low $T$, when the defects are dilute,
and the height field can be constructed locally around each defect.
It then turns out that the height field,
considered on a closed 
loop surrounding a defect, is not a single valued function.\cite{Escher}

The topological charge assignment in this situation
is facilitated by interpreting the defects 
as screw dislocations, centered at the unpaired triangles.
Positions of the dislocations are seen as large ($2\times 2$) triangles in 
Fig.~\ref{snapshots} (F). After 
integrating $\nabla h$ over a contour enclosing a single triangle, 
one obtains a positive or negative 
mismatch in the height $h$ equal to 
the dislocation Burgers vector
\be \label{def-b-KT}
b_{\triangle} = \oint \partial_i h\,dx_i =
\pm 2\ell = \pm 6b \,.
\ee
The topological charge algebra is indeed identical to that
of dislocations (${Z}$).
Following a loop around two dislocations of opposite sign (\ref{def-b-KT}),
gives net zero charge, $\oint \partial_i h \, dx_i=0$.
The topological nature of the defects constrains the dynamics.
During the MC simulation, 
the dislocations originate and disappear only in pairs. 

The minimal energy cost required to
create a defect can be estimated as the energy for unparing 
of two triangles.
Each unpaired triangle adds one more frustrated bond
which can be shared by neighboring triangles
[see Fig.~\ref{snapshots} (F)].
The corresponding energy cost is $2V(a)$ per unpaired triangle.
At sufficiently low temperatures the probability of
creating a pair of defects is therefore exponentially 
suppressed, $P\sim e^{-4V(a)/T}$. 
On the other hand, at a high temperature
$T\gg V(a)$ the defects are so abundant that
the height field ceases to exist even as a local notion.
In this high temperature state, 
marked {\it disordered phase} in Fig.~\ref{phasediagram},
electron hopping is uncorrelated, with
the interaction enforcing single or zero occupancy,
and otherwise playing no role.

At lower temperatures, $T\le V(a)$, 
with the fugacity of a single defect decreasing at least as
$e^{-2V(a)/T}$, the defects quickly become dilute. 
In this case, the height field is defined locally in the entire plane,
except the vicinity of the defects. 
The collective behavior of defects in the correlated phase,
such as the Kosterlitz-Thouless unbinding transition,
is controlled by entropic effects which will be analyzed below 
in Section~\ref{sec:correlated-phase}.

\begin{figure}[b]
\includegraphics[width=2in]{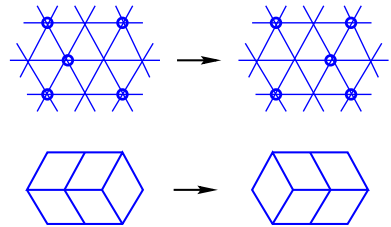} %charge_rhomb.eps
 \caption[]{(Color online) 
The concept of free charges in the correlated phase, 
which can move without changing the number of frustrated bonds, is
illustrated.
{\it Above:} An example of a free charge move;
{\it Below:} The corresponding change in the rhombic tiling.
}
\label{free_charge}
\end{figure}

The charge rearrangements that do not produce or destroy defects cost 
no energy in the \tiafm model limit $d\ll a$. 
In the case of a long range interaction,
such rearrangements generally cost finite 
energy which is determined by the $V_{nnn}$ interaction strength.
An example of a typical single electron move that preserves 
the structure of rhombic tiling is shown in Fig.~\ref{free_charge}.
We find that the state with a height field, despite being constrained by 
the triangle pairing condition, allows for sufficiently large number 
of movable charges which can lead to macroscopic charge rearrangements
which do not produce defect pairs.
The MC conductivity appears to be related to these rearrangements, 
hereafter referred to as \emph{free charges}. 
As Fig.~\ref{free_charge} illustrates, 
the 
moves associated with free charges 
can be interpreted geometrically as changing 
the $3d$-lifted tiling by
adding or removing two adjacent cubes. 
This operation does not introduce defects or discontinuities
in the lifted surface, preserving its 3d continuity.

%%%%%%%%%%%%%%%%%%%%%%%%%%%%%%%%%%%%%%%%%%%%%%%%%%%%%%%%%%%%%%%%%%%%
%%%%%%%%%%%%%%%%%%%%%%%%%%%%%%%%%%%%%%%%%%%%%%%%%%%%%%%%%%%%%%%%%%%%

In our MC study, with electron densities $1/3 \le n \le 2/3$, 
we found that there is a temperature interval in which
the number of topological defects (unpaired triangles) is small,
and thus the height order parameter is well defined. 
Simultaneously, the density $n_f$ of the free charges
(see Fig.~\ref{free_charge}) 
was observed to be large. 
During MC simulations we accumulate a
histogram for $n_f$ and find that this is a nonconserved
quantity with a broad Gaussian distribution. The mean $\overline{n_f}$ is
of the order of $10-50\,\%$ of the total electron density $n$ for
the temperatures 
\be \label{height-fluct}
V_{nnn}\lesssim T\lesssim V_{nn} 
\ee
between the nearest neighbor interaction
$V_{nn}$ which enforces pairing correlations,
and the next-nearest interaction
$V_{nnn}$ which controls freezing at low temperature.

The correlated nature of the charge state at these temperatures,
owing to the continuous height variable constraint, 
leads to a peculiar picture of charge flow. 
The notion of individual electron hopping 
has to be replaced by a more adequate picture involving 
fluctuations of the height field giving rise to charge movement. 
We observe that the Ohmic conductivity
remains finite (Fig.~\ref{condvsT}) in the 
correlated fluid temperature interval,
while the topological defects freeze out. It appears that 
the height fluctuations on their own
are sufficient for charge transport
and conductivity. The elementary height fluctuations,
embodied in the notion of free charges (Fig.~\ref{free_charge}),
represent MC moves of an effective nonconserving (type A) dynamics
of the 2d surface. Based on this idea, 
in Section \ref{sec:roughening}
we shall develop a continual 
approach to describe the dynamics in the correlated state.

One may expect that the free charges, although nominally allowed 
to only move back and forth (Fig.~\ref{free_charge}),
can propagate over the whole system. Microscopically this happens since a move of
one free charge unlocks subsequent moves of other free charges.
This picture is consistent with our observations based on MC dynamics: 
In Sec.~\ref{sec:cond-disl} we find a non-vanishing contribution 
to the conductivity in the absence of dislocations. 
The dislocationless conductivity points to the importance of free charges 
in transport. However, more work will be needed to address other properties 
of free charges dynamics, such as the microscopic transport mechanism and ergodicity.
%the behavior of fluctuations 
%of the height field in the correlated phase 
%studied below in Section~\ref{sec:correlated-phase}. 
%The logarithmic growth (\ref{corr-gaussian}) of the height correlation 
%function with distance, observed in the absense of dislocations,
%points to spatial delocalization of the free charges.

To assess the relevance of this conductivity mechanism
for real systems one needs to understand several issues, 
the most important one being the role of disorder.
Although our MC study of the disordered problem was not too extensive
and, in partricular, restricted to not too low 
temperatures, it allows to draw some assuring conclusions.
In general, for weak disorder we observe no change in
the qualitative features of the dynamics.
We studied the 
dc conductivity in presence of a finite amount of disorder,
modeled by the random potential
$\phi(\vec r_i)$ in the Hamiltonian (\ref{def-Hs}).
The MC simulation used statistically uncorrelated random 
$\phi(\vec r_i)$ with a uniform distribution in the interval 
$-\phi_0 \le \phi(\vec r_i) \le \phi_0$, 
where $\phi_0 \sim V_{nnn}= V(\sqrt{3}a)$. 
The only significant difference observed was 
a relatively more slow time averaging in the presence of disorder,
leading to enhanced fluctuations of the conductivity recorded
as a function of temperature.
Otherwise, the qualitative behavior of the conductivity
was found to be the same as in the clean system.
In particular, the zero bias conductivity remains finite in the correlated
fluid phase in the presence of disorder. 
We attribute the robustness of conductivity against 
weak disorder
to the nonconserving dynamics 
of the height variable, which is not pinned by the disorder 
in the temperature interval of interest.

%%%%%%%%%%%%%%%%%%%%%%%%%%%%%%%%%%%%%%%%%%%%%%%%%%%%%%%%%%%%%%%%%%%%%%%%%%
\section{Dislocation-mediated versus height field-mediated conductivity}
\label{sec:cond-disl}

\nin
Here we investigate in detail the effect of dislocations
on electron transport in the correlated phase.
As noted above, certain charge moves,
associated with ``free charges'' (Fig.\,\ref{free_charge})
have relatively low energy cost.
These moves 
correspond to height field fluctuations which 
do not produce topological defects.
The presence of the pairing correlations described 
by the height variable and
the dynamics of free charges
associated with these correlations, 
poses an interesting question regarding the
relative contribution of the free charges and dislocations to the transport.

\begin{figure}[t]
\includegraphics[width=3in]{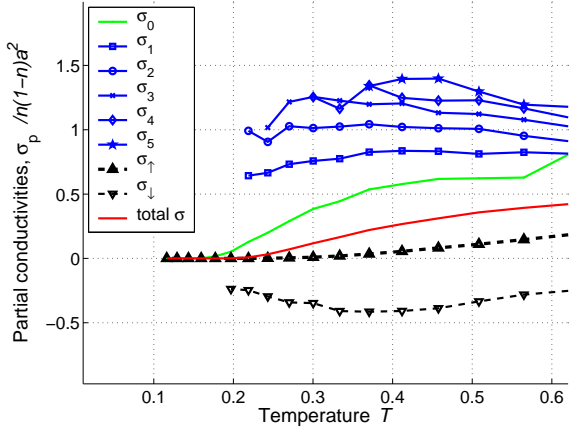} %partial_cond.eps
\caption[]{(Color online)
Partial conductivities $\sigma_p$, $p=0,...,5$, 
which account for MC moves in the presence of $p$ dislocation pairs (see text),
for the 12$\times$12 patch, density $n=\frac12$, screening length $d=2a$,
Eq.\,(\ref{interaction}).
The dislocationless conductivity $\sigma_0$ approaches zero continuously 
at the freezing transition, 
while the conductivities $\sigma_{p}$, $p\neq 0$ remain {\it finite}.
Separately shown are the conductivities $\sigma_{\uparrow}$ and $\sigma_{\downarrow}$
that correspond to 
the MC steps changing the number of dislocation pairs.
(Note the negative sign $\sigma_{\downarrow}<0$.)
}
\label{fig:partial-cond}
\end{figure}

One approach to understand the role of dislocations would be to 
forbid entirely the charge hops that introduce new dislocation pairs 
and study MC conductivity in such a system. We have tried to modify
the MC simulation in this way, and found that the dislocationless
system exhibits
finite conductivity solely due to free charges, indicating
that the dislocations are not essential for conductivity.
This method, however, does not allow to compare 
the role of dislocations and free charges quantitatively, 
since the rule introduced to eliminate dislocations 
alters the dynamics in an uncontrollable way. In fact, since 
the definition of conductivity changes in the dislocationless MC, although 
we indeed observe a nonzero conductivity, it is difficult to compare
the resultant ``dislocationless'' conductivity with that 
of the original problem.
Instead, we adopt a different, more gentle approach.
We employ the same dynamics as above
(conserving, or type B),
in which we detect dislocations
and 
analyze {\it partial conductivities} $\sigma_{p}$
in the presence of $p=0,1,2,..$ dislocation pairs
(Fig.~\ref{fig:partial-cond}).

%%%%%%%%%%%%%%%%%%%%%%%%%%%%%%%%%%%%%%%%%%%%%%%%%%%%%%%%%%%%%%%%%%%%
%%%%%%%%%%%%%%%%%%%%%%%%%%%%%%%%%%%%%%%%%%%%%%%%%%%%%%%%%%%%%%%%%%%%
%%%%%%%%%%%%%%%%%%%%%%%%%%%%%%%%%%%%%%%%%%%%%%%%%%%%%%%%%%%%%%%%%%%%
\begin{figure}[t]
\includegraphics[width=3in]{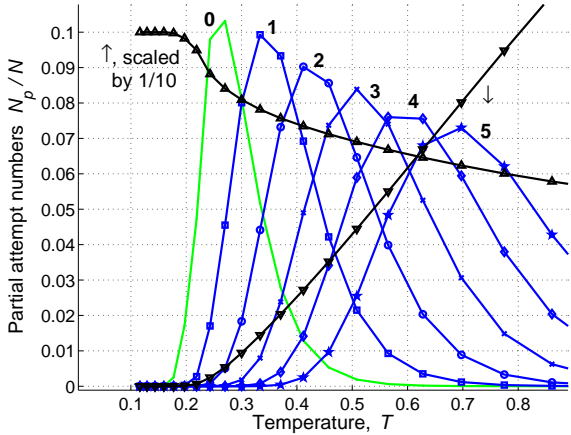} %attempts.eps
\caption[]{(Color online)
Partial attempt numbers scaled by $\N$ define
normalized weights $\N_p/\N$, $p=$0,...,5, $\uparrow$, $\downarrow$, 
for the partial conductivities $\sigma_p$ of Fig.~\ref{fig:partial-cond}.
The total $\N$ [Eq.~(\ref{attempts-normalization})]
was approximately constant, varying by less
than $5\%$
in the whole temperature range.
With the $\N\approx 3\cdot 10^8$ per
temperature step, the whole simulation took about 15 
hours on a 1.25 GHz PowerPC.
}
\label{fig:attempts}
\end{figure}
%%%%%%%%%%%%%%%%%%%%%%%%%%%%%%%%%%%%%%%%%%%%%%%%%%%%%%%%%%%%%%%%%%%%

Let us outline the corresponding MC procedure.
We introduce a small bias ${\bf E}$ across the finite system
with periodic boundary conditions,
and define the partial conductivities $\sigma_{p}$ 
similarly to the total conductivity
calculated as described in Section~\ref{sec:conductivity}, Eq.~(\ref{def-cond}):
\be \label{def-partial-cond}
{\bf j}_{p} = \sigma_{p} {\bf E} .
\ee 
The partial currents $j_{p}$ at a fixed number $p$ of the dislocation pairs 
are defined similarly to the total current, 
Eq.~(\ref{def-j-cond}), with 
the numbers of hops $\N_{\pm}$, $\N=\N_++\N_-$, replaced
by the $p$-dependent ${\N_{p}}_{\pm}$ and $\N_{p}={\N_{p}}_++{\N_{p}}_-$.
Here $\N_{p}$ is the total number of MC \emph{trials}, or attempted
hops that may or may not lead 
to real hops, in which a pair of randomly selected 
neighboring sites is such that the charge hop between them
would preserve the number $p$ of dislocation pairs.
Accordingly, ${\N_{p}}_+$ and ${\N_{p}}_-$ descibe
charge hop trials along and opposite to the applied field
with a fixed number $p$ of dislocations.
The actual hop is then attempted with a Boltzmann probability 
(\ref{probabilitiesB}), as discussed in Section~\ref{sec:themodel}.
In our simulation we studied the system 
$12\times 12$ at density $n=\frac12$.

The partial conductivities $\sigma_p$, $p=0,1,2,...$,
describing the contribution to transport of charges hopping
in the presence of the $2p$ dislocations,
need to be complemented by two additional parts,
$\sigma_{\uparrow}$ and $\sigma_{\downarrow}$, describing
the conduction processes which alter the 
number of dislocations.
Indeed, a significant fraction of the occupied-empty 
site pairs chosen in the MC simulation are such that upon
a charge hop between them 
the dislocation number
would either increase or decrease.
Such hops, if selected by the Boltzmann
probabilities, can also carry current in the presence of the applied bias
(as illustrated in Fig.~\ref{fig:partial-cond}).
With the corresponding numbers of the MC trials denoted by $\N_{\uparrow}$ and 
$\N_{\downarrow}$, the total number of attempts is
\be \label{attempts-normalization}
\N = \N_0 + \N_1 + \N_2 + ... + \N_{\uparrow} + \N_{\downarrow} .
\ee
The partial MC attempt numbers $\N_p$, $\N_{\uparrow}$, $\N_{\downarrow}$ 
temperature dependence is summarized 
in Fig.~\ref{fig:attempts}. 

As Figs.~\ref{fig:partial-cond} and \ref{fig:attempts} illustrate, 
the average number of dislocations in the $12\times 12$ patch
evolves from about 5 at $T\approx 0.7 V(a)$ to zero in the vicinity of the 
freezing transition.
As the temperature lowers, most of the time the system attempts to create 
dislocation pairs ($\N$ is dominated by $\N_{\uparrow}$),
however, almost all of these 
attempts are discarded due to their exponentially low Boltzmann weight,
resulting in the negligible current $j_{\uparrow}$ and partial conductivity
$\sigma_{\uparrow}$ (see Fig.~\ref{fig:partial-cond}).

We note a drastic difference between the dislocationless conductivity
$\sigma_0$ and the partial conductivities $\sigma_{p\neq 0}$ near the freezing
transition at $T_c\approx 0.2 V(a)$.
Whereas the former approaches zero continuously
as $T\to T_c$, the latter have an apparent 
step-like discontinuity. 
However, the contributions of partial conductivities $\sigma_p$, $p\neq 0$,
to the total conductivity 
\be \label{cond-total}
\sigma=\N^{-1}\sum_{p=0,1,2,..., \uparrow,\downarrow} \N_{p} \sigma_{p} 
\ee
are small,
since 
the weights $\N_{p\ne0}/\N$ drop very quickly as $T$ approaches $T_c$
(Fig.\,\ref{fig:attempts}).
In a relatively small $12\times12$ system studied here,
the dislocations are almost always totally absent near $T_c$.
In this case
the total conductivity (\ref{cond-total}) is dominated by $\sigma_0$
(Fig.~\ref{fig:partial-cond}), due to  
small $\N_{p\neq 0}/\N$.
This is consistent with the observation that
the total conductivity decreases to zero in a continuous fashion
near $T_c$, similar to $\sigma_0$,
with the step-like contribution of $\sigma_{p\ne0}$  
being inessential.

One cannot exclude, however, a different regime in a large system,
where the discontinuous part $\sigma_{p\ne0}$
of the total conductivity (\ref{cond-total}) 
can become dominant in the thermodynamic limit, 
due to dislocation number growing with system size.
Even in this case, we expect the dislocationless 
contribution to the conductivity to remain significant.
In that regard, we note an approximately constant increment 
in conductivity
$\Delta_p\sigma=\sigma_{p+1} -\sigma_p \approx {\rm const}$ 
when the number of dislocations 
increases by one, $p\to p+1$, observed at temperatures 
not too close to freezing (Fig.~\ref{fig:partial-cond}).
The approximately $p$-independent $\Delta_p\sigma$ provides an estimate 
of the conductivity per dislocation, suggesting that
in the dilute regime the dislocations
contribute to transport approximately independently.
However, the dislocationless part $\sigma_0$ is a few times larger
than $\Delta_p\sigma$, providing a constant offset to the 
$\sigma_p$ vs. $p$ dependence. It is thus not unconceivable
that, if the dislocations are suppressed, e.g. due to Kosterlitz-Thouless
transition, or otherwise, the conductivity would remain finite, 
dominated by the dislocationless contribution. 

This conclusion is reinforced by the results of MC simulation
with the rules modified so that the dislocations are totally excluded, 
as discussed above. In this case, we find finite conductvity, 
behaving as a function of temperture at $T>T_c$ in a way similar to $\sigma_0$. 
Interestingly, the freezing transition is unaffected by this alteration,
with the same value of $T_c$ observed under changed MC rules.
These observations, in our view, leave no doubt that the conductivity
does not entirely depend on dislocations. 
The transport of free charges, responsible for the 
dislocationless contribution $\sigma_0$, gives rise to finite conductivity
independent of dslocations.

Interestingly, the partial conductivity due to annihilation of the topological
defect pairs is negative, $\sigma_{\downarrow}<0$. While we do not have
a complete understanding of this observation, 
we mention one possible explanation 
which involves a nonlinear effect. We suppose that a large enough
electric field can perturb the system  
so that its relaxation back to equilibrium will be accompanied by 
release of a charge in the opposite direction. However, our MC study
indicates that
the relative weight 
of such processes, and thus its contribution to the 
conductivity, is small near the freezing transition (Fig.\,\ref{fig:attempts}).
Therefore, in our view, 
the associated nonlinear conductivity is not essential for
the interpretation of the numerical data.

The described analysis of partial conductivities does not
reveal the microscopic mechanism by which dislocations facilitate
transport. One possibility is that the conduction is
due to the motion of  
dislocations themselves (since they are charged objects).
Another possibility is that the mere presence of  
dislocations
facilitates the hops of free charges in their vicinity and results 
in a finite conductivity. 
While our MC study does not provide a definite answer,
it is possible that the two alternatives 
are not unrelated, since the motion of a dislocation
can happen due to a free charge movement right next to it.

To summarize, we have shown that the dislocation pairs facilitate conductivity,
and have compared their contribution with dislocationless conductivity
due to free charges. While in the model analyzed here there is no 
obvious way to completely separate these contributions, 
due to the absence of a dislocation binding
transition, we conclude that both contributions are present,
with their relative importance depending on the dislocation density.

%%%%%%%%%%%%%%%%%%%%%%%%%%%%%%%%%%%%%%%%%%%%%%%%%%%%%%%%%%%%%%%%%%%%%%%%%%
\section{Fluctuations in the correlated phase: Dislocations unbinding}
\label{sec:correlated-phase}

The height field describes the effect of frustration on charge dynamics 
by providing a nonlocal change of variables that helps to keep track of 
the local correlations between charges. However, as we found above, in general
the height variable cannot be globally defined due to the 
presence of dislocations. Nonetheless, at relatively low temperature,
when dislocations are dilute, the 
height field provides a useful description on a local level. 

Here we adopt the view that the dynamics of the height field is simpler than 
the underlying charge dynamics.
Indeed, while the latter is strongly constrained, the fluctuations
of the height field appear to be Gaussian. The height variable
also provides a natural continual description employed to 
study dislocation unbinding (this Section),
roughening transition (Section~\ref{sec:roughening}),
the dynamics (Section~\ref{sec:dynamics}) as well as the 
freezing into commensurate states (Sec.~\ref{sec:freezing} above).

The MC dynamics in the correlated state
can be interpreted as the height field fluctuations. 
These fluctuations dominate  
in the temperature interval (\ref{height-fluct})
where the dislocations are dilute. 
Thus here we consider the height field ignoring 
for some time the effect of dislocations.
We study the height fluctuations numerically and find that
in the continuum limit $r\gg a$ they
are described by the partition function of the form
\be \label{Z-gaussian}
Z_0 = \int \! {\cal D}h(\vec{r}) \; 
e^{- \int \! d^2\vec{r} \; {\kappa \over 2 } (\nabla h)^2} \,,
\ee
where $\kappa$ is an effective stiffness of the height surface,
with temperature incorporated into $\kappa$.

The Gaussian partition function (\ref{Z-gaussian}) 
has been introduced by Bl\"ote and Hilhorst,\cite{Blote82} 
and by Niehnuis et al.\cite{Blote84} for 
the height field in the \tiafm model at $T\to 0$.
The corresponding stiffness  
has been obtained from the exact solution\cite{IsingAFM,Blote84}:
\be\label{kappa-tiafm}
{\kappa_{\triangle\rm IAFM} } = {\pi \over 9b^2} \,. 
\ee 
[Note that, instead of the stiffness $\kappa$, 
Ref.~\onlinecite{Blote84} 
employs the ``renormalized temperature'' $T_R=2\pi/\kappa b^2$,
with $T_R=18$ for the \tiafm.]
The equilibrium height fluctuations have been 
studied numerically\cite{higher-spin-AFM,chakraborty}
in the case of the nearest neighbor interactions,
confirming the result (\ref{kappa-tiafm}) for the \tiafm.

Here we extend these results to the model with long range interaction.
We study the fluctuations of the height field numerically, by 
assigning the heights $h(\r_i)$ to the triangular lattice
sites $\r_i$ according to the procedure
described in Appendix. The results of this study can be summarized
as follows.

(i)
At fixed temperature,
we accumulate the
histograms $\P[h(\r_i)]$ of height values for 
a set of points $\{\r_i\}$  in the $N\times N$ patch.
We find that the fluctuations are Gaussian,
\be \label{histogram-gaussian}
\log \P[h(\r_i)] \propto -h^2(\r_i) \,,
\ee
independently for each point $\r_i$. This suggests that
a partition function of the form (\ref{Z-gaussian})
can be employed.

(ii)
To calculate the effective stiffness 
for a system with the long-ranged interaction
(\ref{interaction}), we fit the two-point height correlator 
to the one following from Eq.~(\ref{Z-gaussian}): 
\be \label{corr-gaussian}
\left< (h(\vec{r}_i)-h(\vec{r}_j))^2\right> = 
{1 \over \pi \kappa} \ln {|\vec{r}_{ij}| \over r_0} \,,
\ee
with the length $r_0$ of the order of the lattice constant $a$. 
For this analysis we generate random MC configurations, 
selecting the height surfaces of zero average slope, 
defined as in Eq.~(\ref{def-slope}) above.

We tested this procedure by calculating the stiffness 
in the nearest neighbor interaction limit $d\ll a$ at $n=1/2$.
The MC dynamics recovers the value $\kappa_{\triangle\rm IAFM}$,
Eq.~(\ref{kappa-tiafm}), within a few percent accuracy.
With the long range interaction,
the relationship (\ref{corr-gaussian}) still holds, allowing us to 
determine $\kappa$. 
We systematically find the values 
of $\kappa$ below $\kappa_{\triangle{\rm IAFM}}$,  
Eq.~(\ref{kappa-tiafm}).

(iii) 
The measured stiffness $\kappa$ can be used to assess the possibility
of a dislocation-binding transition of the Kosterlitz-Thouless kind.\cite{KT,KTHNY} 
If exist, such a transition between the disordered and
correlated fluid phases would realize\cite{Blote91,chandra-coleman-ioffe} 
the KTHNY 2d melting scenario. 
Dislocation binding takes place above the universal threshold value
\be \label{kappa-KT}
\kappa > 
\kappa_{\rm KT} = {8\pi \over b_{\triangle}^2} \,,
\ee
with $b_{\triangle}$ the Burgers vector of the dislocation. 
Using the value from Eq.~(\ref{def-b-KT}), 
one obtains $\kappa_{\rm KT} = 2\kappa_{\triangle\rm IAFM}$. 
Thus the KT transition is absent in the zero-field \tiafm.\cite{Blote84} 

We find that this conclusion remains valid for the 
long range interaction.
For the interaction (\ref{interaction}), 
the stiffness {\it decreases} as a function of $d$. 
For typical $d$ used in this work, stffness values are 
a few times smaller than $\kappa_{\triangle\rm IAFM}$. 
Although such a behavior may seem counterintuitive
(one could expect the system to become stiffer for an interaction of  
longer range), it is in fact consistent with the scaling arguments presented 
below in Sec.~\ref{subsec:ren-scaling}. 
It is also compatible with the observation\cite{Blote84}  
that the ferromagnetic second-neighbor coupling
causes increase of stiffness. 
In agreement with this, 
adding longer-range antiferromagnetic couplings should 
lead to the decrease of $\kappa$.

We conclude that the dislocations are always unbound in our system.
This is indicated by a crossover (rather than a sharp transition)
between the disordered and correlated phases 
in the phase diagram, Fig.~\ref{phasediagram}.
Thus, in the thermodynamic limit, the dislocation pairs are present at any
temperature.
While their concentration may be small, it is nonzero,
and, strictly speaking, the height field
is not well-defined in an infinite system. 
However, in a finite system
we often observe that dislocations are absent at low temperature, 
which justifies the height field as a local notion.

%%%%%%%%%%%%%%%%%%%%%%%%%%%%%%%%%%%%%%%%%%%%%%%%%%%%%%%%%%%%%%%%%%%%%%%%%%
\section{The role of height fluctuations near freezing transition}
\label{sec:roughening}

%% Korshunov says:
%{\small 
%It seems to me that after You have demonstrated in Sec. VII that for
%the interaction (2) the value of kappa is even lower than in the case of
%only nearest-neighbor interaction, the possibility of having a roughening
%transition (into the flat state with zero slope) is already closed.
%According to Nienhuis et al. the roughening transition can happen only
%when the value of kappa is 9/2 times higher that the value at which
%the dislocations become coupled, and your kappa is always much smaller.
%Thus, although the analysis of Sec. VIII is certainly useful in other
%respects, its opening statement that large fluctuations of height may lead
%to the freezing into the flat state looks to me misleading.
%}

Although the height surface fluctuates freely in the correlated phase, 
the stiffness values $\kappa$ obtained from the MC dynamics are found to 
be below the bound\cite{Blote84} necessary for 
the fluctuation-induced roughening transition.\cite{roughening}
In this respect, the case of the long-range interaction 
(\ref{interaction}) appears to be qualitatively similar 
to that of the \tiafm in the small field.\cite{Blote84,Blote91}
The purpose of the present Section is to rationalize this 
similarity, and quantify the reduction 
of stiffness (as compared to that of the \tiafm) by employing 
renormalization group methods.

%At low temperature the height fluctuations may become 
%non-gaussian and lead to freezing. 
%Here we analyze the possibility of a
%{\it roughening transition}\cite{roughening}
%of the height field,
%as a mechanism for freezing into a solid phase.
%Such a transition would be of an infinite order, described by 
%the Kosterlitz-Thouless universality,
%albeit different from the dislocation-binding KT transition
%considered above.

Below we construct the free energy for 
the charge denstity and the height field, 
with a coupling of a sine-Gordon form, and present its scaling analysis.
We obtain a bound on the effective stiffness $\kappa$,
Eq.~(\ref{kappa-compare}), and, by comparing
it to that found from the MC dynamics, 
determine that the continuous roughening 
transition does not take place.
Rather, as discussed in Section~\ref{sec:phase-diagram}, 
freezing occurs via a finite order transition. 
The bound (\ref{kappa-compare}) indicates that the  height fluctuations
are irrelevant, and, as a result, 
allows us to rule out both instances of the continuous 
transitions (dislocation-unbinding, and roughening).
%of the Kosterlitz-Thouless universality.
%also allows us to relate the roughening 
%and dislocation-binding KT transitions, 
%and to understand why the absence of 
%the former implies the absence of the latter.
Subsequently, in Section~\ref{sec:dynamics}, we use the developed formalism
to study
%the effect of the height fluctuations on the 
equilibrium current fluctuations and conductivity.

%%%%%%%%%%%%%%%%%%%%%%%%%%%%%%%%%%%%%%%%%%%%%%%%%%%%%%%%%%%%%%%%%%%%%%%
\subsection{Free energy}
\label{sec:free-energy}

\nin 
The correlated state is described
in terms of the height variable $h(\r)$ and charge density $n(\r)$.
Microscopically the state of the system is defined uniquely by
specifying either the former or the latter.
However, 
here we demonstrate by scaling analysis that 
at large length scales the height and density fluctuations decouple.
This obeservation allows us to employ a partition function
with $h$ and $n$ as independent variables,
\be \label{Z-correlated}
Z_{\rm corr} = \int \! {\cal D}h(\r) {\cal D}n(\r) \, 
e^{-\F[h, n]} \,.
\ee
Here the free energy 
\be
\label{F-tot}
\F[h, n] = \F_h + \F_n + \F_{\rm int} 
\ee
is a sum of the contributions of the height field, the charge density, and 
their interaction.
As before, the temperature is incorporated into $\F$.

The phenomenological free energy $\F_h$ has the form
\be \label{F-h}
\F_h = \int \! d^2\r  
\lp {\kappa\over 2}(\nabla_{\r} h)^2 
+g \cos {2\pi h\over b} 
+ f \cos {\pi h\over b} \rp , 
\ee
with $\kappa$ the stiffness.
Here the term $\cos {2\pi h\over b}$ assigns higher statistical weight to the field
configurations that pass through the points of the 3d cubic
lattice, with the period $b$ in height given by Eq.~(\ref{def-b}).
The third term in (\ref{F-h}) describes the coupling 
of charge to the chemical potential, in our case realized as the gate voltage, 
\be \label{f=Vg}
f = C \bar n  e\Vg \,,
\ee
with $\bar n$ the average charge density and $C\simeq T^{-1}$.
The coupling of the form $\cos {\pi h\over b}$, 
with the period $2b$ in height, arises because
opposite charges occupy two
different sublattices of the 3d lattice, alternating in the 
height direction.\cite{Blote84} 
The third term in Eq.(\ref{F-h}) is always more relevant than the second one
in the sense of scaling.

Consider now the charge density $n_{\r}$.
Electrons interact with each other and with an external electrostatic
potential $\Phi^{\rm ext}(\r)$, which gives 
\be 
\F_n = {1\over 2}\, \int \! d^2\r \, d^2\r' \, 
n_{\r} U_{\r-\r'} n_{\r'} 
+ \int \! d^2\r \, n_{\r} \Phi^{\rm ext}(\r) 
\label{F-n}
\ee 
with the interaction
$U_{\r-\r'}$ of the form (\ref{interaction}).

The coupling between the density and the height variable should be periodic 
in $2b$ for the same reason as in Eqs.~(\ref{F-h}) and (\ref{f=Vg}),
with $\bar n$ replaced by the fluctuating density:
\be \label{F-int}
\F_{\rm int} = \lambda \, \int \! d^2\r \; n_{\r} 
\; \cos {\pi h_{\r}\over b} \,.
\ee
Scaling analysis of the problem (\ref{F-tot}) is presented below.

%%%%%%%%%%%%%%%%%%%%%%%%%%%%%%%%%%%%%%%%%%%%%%%%%%%%%%%%%%%%%%%%%%%%%%%%%%
\subsection{Scaling analysis}
\label{subsec:ren-scaling}

\nin
The nonlinear terms in Eqs.~(\ref{F-h}), (\ref{F-int}) acquire nontrivial
scaling dimensions due to fluctuations. 
At one loop, the anomalous dimensions of the couplings $f$, $\lambda$ and
$g$ are 
\bea \label{scaling-f}
X_f = X_{\lambda} = {2\kappa^*\over \kappa} , \\
\label{scaling-g}
X_g = {8\kappa^*\over \kappa} ,
\eea
where we define
\be \label{def-kappa-star}
\kappa^* \equiv {\pi \over 8 b^2} \,.
\ee
The term $f\cos {\pi h\over b}$ becomes relevant when 
$X_f < 2$, or 
\be \label{fixed-f}
\kappa > \kappa_f = \kappa^* \,.
\ee
The coupling $g$ becomes relevant at a larger stiffness: 
\be \label{fixed-g}
\kappa > \kappa_g = 4\kappa^*  \,.
\ee
The coupling $\lambda$ between the
density and height fields is relevant when 
the scaling dimension of the gradient term is larger than
the sum of scaling dimensions of $\lambda$ and $n$.
The field $n$ is not renormalized since its free energy is quadratic.
From (\ref{F-n}), the bare dimension of $n$ is $X_n=3/2$.
Hence the height-density interaction is relevant when 
\be \label{fixed-lambda}
X_{\lambda} + X_n < 2 \,, \quad {\rm or} \quad
\kappa > \kappa_{\lambda} = 4\kappa^* \,.
\ee

The nonlinear terms in (\ref{F-h}) do not renormalize the stiffness $\kappa$
at one loop. However, the stiffness may obtain a one loop correction
$\delta \kappa$
due to the interaction with the density, Eq.~(\ref{F-int}). 
Below we consider such a possibility.

Since the free energy $\F_n$ is Gaussian, 
we integrate out the field $n_{\r}$ in (\ref{Z-correlated}) 
and obtain an effective free energy for the height field:
\be \label{tilde-F-h}
\tilde{\F}_h = \F_h - {\lambda^2\over 2}\, 
\sum_{\k} \lp \cos{\pi h\over b}\rp_{-\k} \, {1\over U_{\k}} \, 
\lp \cos{\pi h\over b}\rp_{\k} \,.
\ee
Consider the case of the screened Coulomb interaction
of the form (\ref{interaction}),
\be \label{U-k}
U_{\k} = {2\pi\over k}\, \lp 1 - e^{-kd}\rp .
\ee
At length scales larger than the screening length, $kd \ll 1$, 
\be \label{AplusBk}
{1\over U_{\k}} \approx {1\over 2\pi} \, \lp A + Bk \rp 
\ee
with  $A=1/d $, $B = 1/2$.
For the unscreened interaction, 
$(U_{\k})^{-1}$ is of the form identical to  
(\ref{AplusBk}) with $A=0 $, $B = 1$.
The local term $A/2\pi$ corrects the coupling
constant $g$ and does not contribute to the stiffness.
The correction to stiffness $\delta \kappa$ arises from the nonlocal
part of (\ref{tilde-F-h}) with $U_{\k}^{-1} \to  Bk/2\pi$.
Expanding $\cos\pi h/b = \lp e^{i\pi h/b} + e^{-i\pi h/b}\rp/2$,
we note that 
the terms $(e^{i\pi h/b})_{-\k} (e^{i\pi h/b})_{\k} + {\rm c.c.}$
have a larger scaling dimension than the cross terms,
$(e^{-i\pi h/b})_{-\k} (e^{i\pi h/b})_{\k} + {\rm c.c.}$ .
Neglecting the former and expanding the latter, 
we arrive at the following stiffness correction:
\be \label{ren-kappa}
\delta\F = -\frac12 \sum_\k J_k \kappa^* k^2 h_{-\k} h_\k
%\delta \kappa = -  J \kappa^* ,
\ee
where we introduced a nonlocal coupling 
$J_k \equiv 2B \lambda^2 /|\k| $.
Note that the screening length $d$ does not contribute to
the stiffness correction, affecting only
the value of $B$.
The negative sign of the stiffness correction (\ref{ren-kappa})
is consistent with the observed stiffness reduction in MC simulation with 
long range interaction, compared to $\kappa_{\triangle {\rm IAFM}}$.

The coupling $J$ is renormalized via integrating out the
height fluctuations $h_{\k}$ with 
$1/l_1 < k < 1/l_0$:
\be \label{ren-J}
J_{l_1^{-1}} = J_{l_0^{-1}} \, 
\lp {l_1 \over l_0}\rp^{1-4\kappa^* / \kappa} \,.
\ee
The stiffness correction (\ref{ren-kappa}) is relevant
only when the condition (\ref{fixed-lambda}) holds.
Therefore, there is no qualitative effect on the Gaussian
behavior (\ref{Z-gaussian}).

We point out that the negative contribution to stiffness, Eq.~(\ref{ren-kappa}),
can be understood as an effect of long range interaction as follows.
In the 3d cubic crystal picture, the charges of plus and minus sign reside on 
the even and odd sublattices. With respect to the (111) direction used to introduce
the height variable, these sublattices define alternating stacks of planes normal
to the (111) axis, each filled with charges of one sign. Now, every surface with
a small gradient of the height field can be associated with a set of terraces
of alternating total charge and of width inversely proportional to the gradient
of the height field. Since the long range Coulomb interaction favors the states
with plus and minus charges well-mixed, one expects it to favor more narrow terraces
compared to wider terraces. This means that the long range repulsive interaction
contribution to the energetics of the height field must be negative of 
$(\nabla h)^2$, so that the favored states have the maximal possible height gradient.
This observation is consistent with both the negative sign of the result 
(\ref{ren-kappa}), and with the stripe-like character of the ground state at 
$n=1/2$ [Fig.~\ref{snapshots}(C), single domain], stabilized by the 
long-range repulsion.

%%%%%%%%%%%%%%%%%%%%%%%%%%%%%%%%%%%%%%%%%%%%%%%%%%%%%%%%%%%%%%%%%%%%%%%%%%%%%%%%
\subsection{Bounds on stiffness}
\label{sec:bounds}

\nin
To summarize the above results, 
%of Sec.\ref{subsec:ren-scaling},
the scaling relations set bounds on the 
effective stiffness $\kappa$ in the freely fluctuating
height model, Eq.~(\ref{Z-gaussian}).
The Gaussian behavior is stable when the nonlinear terms 
are irrelevant, i.e. when
\be \label{kappa-compare}
\kappa < \kappa_f < \kappa_g = \kappa_{\lambda} \,.
\ee
As the temperature decreases, the height surface becomes more rigid.
There are several possible scenarios for a transition from a high-temperature
(rough) to a low-temperature (smooth) phase.
One is the fluctuation-driven roughening transition, involving
the system freezing into a smooth phase 
due to the last term in (\ref{F-h})
when $\kappa$ approaches $\kappa_f$ from below.\cite{roughening,Blote84}

Based on the evidence accumulated in our numerical simulations
we conclude that such a behavior does not occur.
We observe that the measured stiffness $\kappa$
obeys the condition (\ref{kappa-compare}) as long as the height surface 
fluctuates appreciably. 
Hence, in particular, the coupling (\ref{f=Vg}) to the gate voltage is irrelevant 
in the correlated phase of the model with the long range interaction 
(\ref{interaction}).
This conclusion generalizes the observation\cite{Blote84}
that the coupling to external field is irrelevant in the \tiafm
(for sufficiently small field).

Lowering the temperature further produces a 
sharp freezing into a ground state in which the height surface does not fluctuate.
While the freezing is accompanied by a steep increase of $\kappa$ 
to immeasurably high values, the latter takes place at $T\approx T_c$.
Due to narrow temperature interval in which this increase occurs, it is 
difficult to attribute it to the effect of height fluctuations. 
Instead, the observed  
behavior is consistent with an ordinary freezing by a first or second order
phase transition (as discussed in Section~\ref{sec:phase-diagram}).

Above the freezing temperature, in the correlated phase 
the density and height fields effectively decouple, since the coupling
$\lambda$ is irrelevant.  
Indeed, the inequality (\ref{fixed-lambda}) 
cannot be satisfied in the correlated phase, Eq.~(\ref{kappa-compare}). 
This justifies \emph{a posteriori} our phenomenological approach 
of including both the height and the density 
in the partition function (\ref{Z-correlated}).
Similar arguments also explain why 
the negative correction (\ref{ren-kappa}) to the stiffness 
does not cause an instability. The latter would require
$\kappa \geq \kappa_{\lambda}$, the condition forbidden in the correlated phase
by Eq.~(\ref{kappa-compare}).

Finally, the absence of the dislocation-unbinding phase transition
between the correlated fluid and the high
temperature disordered phase (studied in Section~\ref{sec:correlated-phase} above)
can also be understood based on the condition (\ref{kappa-compare}). 
Since $\kappa < \kappa_f < \kappa_{\rm KT}$,
the lower bound (\ref{kappa-KT}) is never reached.
[Similar argument was used in Ref.~\onlinecite{Blote91}
for the \tiafm in zero or small external field].

%%%%%%%%%%%%%%%%%%%%%%%%%%%%%%%%%%%%%%%%%%%%%%%%%%%%%%%%%%%%%%%%%%%%%%%%%%
\section{Dynamics of the height field}
\label{sec:dynamics}

\nin 
The weakly coupled fluctuations of height and density
in the correlated phase lead to interesting dynamical effects.
Employing the Langevin dynamics, here 
we discuss how the interplay between the $h$ and $n$ fields affects the
dynamical properties.
We find the corrections to the
conductivity and compressibility due to the height fluctuations,
perturbative in the height-density coupling $\lambda$.

The non-conserving Langevin dynamics for the height field has the form
\be \label{def-langevin-h}
\partial_t h(\r, t) = - \eta T {\delta \F \over \delta h} + \xi_{\r, t} ,
\ee
with $\F$ given by Eq.~(\ref{F-tot}),  
and $\xi$ the stochastic force, 
\be \label{xi-correlations}
\la \xi_{\r, t} \xi_{\r', t'} \ra 
= \la \xi^2 \ra \, \delta(t-t') \, \delta(\r-\r') \,.
\ee
Since the temperature $T$ is included in $\F$, it multiplies 
the kinetic coefficient $\eta$ in Eq.~(\ref{def-langevin-h}),
so that the form of the fluctuation-dissipation relationship is preserved:
\be \label{einstein}
\la \xi^2 \ra  = 2\eta T \,.
\ee 
The conserving Langevin dynamics for the charge density $n$ is defined using 
the continuity equation 
\be 
\partial_t n + \nabla_{\r} \j = 0 \,, \quad 
\j = - \sigma T \, \nabla_{\r} \, {\delta \F \over \delta n} + \j^L ,
\label{def-langevin-n}
\ee
with the fluctuating extraneous part $\j^L$ obeying
\be
\label{j-correlations}
\la j^L_{\mu}(\r, t) j^L_{\nu}(\r', t') \ra 
= \la (\j^L)^2 \ra \, \delta_{\mu \nu} \, \delta(t-t') \, \delta(\r-\r') \,.
\ee
The conductivity $\sigma$ is related to $\j^L$ variance by the Nyquist formula 
\be \label{nyquist}
\la (\j^L)^2 \ra  = 2\sigma T \,.
\ee
Eqs.~(\ref{def-langevin-h}, \ref{def-langevin-n}) and 
(\ref{F-h}, \ref{F-n}, \ref{F-int}) yield 
\bea \nonumber
\partial_t h(\r, t) &=&  \eta T \lp
\kappa \nabla^2 h  
+ {2\pi g\over b}\, \sin {2\pi h\over b} \right.\\ 
&& \left. + {\pi  (\lambda n + f)\over b}\, \sin {\pi h\over b} \rp
 + \xi(\r, t) \,, \quad
\label{langevin-h}
\eea
\bea \nonumber
\partial_t n(\r, \, t) & = & \sigma T \, \nabla_{\r}^2 
\lp \int \! d^2\r' \, U_{\r-\r'} n_{\r'} + \Phi^{\rm ext}_{\r} \rp  \\
&& - \nabla_{\r} \lp \j^h + \j^L \rp , 
\label{langevin-n}
\eea
where we write the stochastic contribution to the current due to the height
field fluctuations in the form
\be \label{j-h}
\j^h(\r, \, t) = -\lambda \, \sigma T \, \nabla_{\r} 
\cos {\pi h_{\r}\over b}  \,.
\ee

The height field fluctuations provide an additional 
contribution $\delta \sigma^{h}$ to the conductivity, which is 
determined by the fluctuation-dissipation theorem:
\be \label{fdt-wk}
\la j_{\mu}^h(-\omega, -\k) \, j_{\nu}^h(\omega, \k) \ra  
= 2\, \delta\sigma^{h}_{\mu \nu}(\omega, \k) T .
\ee 
The contribution to the conductivity 
that arises from the height field fluctuations is purely longitudinal:
\be \label{def-deltasigma}
\delta\sigma^{h}_{\mu \nu}(\omega, \k) = {\lambda^2 T\over 2} 
\sigma^l(-\omega, -\k) \, \sigma^l(\omega, \k) 
k_{\mu}k_{\nu} \, \tilde{F}(\omega, \k) \,. 
\ee
Here $\sigma^l(\omega, \k)$ is the longitudinal part of the total
conductivity $\sigma$, and $\tilde{F}(\omega, \k)$
is a Fourier transform of the correlator 
\bea \nonumber
F(\r, t) &=& \la \cos {\pi h_{\r,t}\over b} 
\cos {\pi h_{\vec{0}, 0}\over b} \ra_0 \\ 
\label{def-coscos}
&=& {1\over 2} \exp \lf -{\pi^2\over 2 b^2}
\la \lp h_{\r,t} - h_{\vec{0},0}\rp^2 \ra_0 \rf . 
\eea
Averaging in (\ref{def-coscos})
with respect to the Gaussian free energy, $\lambda=g=0$
in Eq.~(\ref{langevin-h}), yields
\bea \nonumber
\la \lp h_{\r,t} - h_{\vec{0},0}\rp^2 \ra_0 
= \sum_{\k, \omega} \frac{\la \xi^2 \ra}{(\eta T \kappa k^2)^2 + \omega^2}
 \left| 1 - e^{i\k\r -i\omega t}\right|^2  \\
= {1\over \pi\kappa} \int_0^{1/a} \!\!\!\! dk \, 
\frac{1-\, e^{-\eta\kappa k^2 |t|}\, J_0(kr)}{k} \,. \quad
\label{h-h-1}
\eea
We are interested in the asymptotic behavior of the height correlator 
at large separation $r\gg a$. 
From the small $k$ expansion under the integral (\ref{h-h-1}) we have
\be
1-e^{-\eta\kappa k^2 |t|}\, J_0(kr) 
\sim k^2 \, \lp {r^2\over 4} + \eta\kappa |t|\rp \,.
\ee
Therefore the height correlator (\ref{h-h-1}) 
asymptotic behavior at $r\gg a$ is consistent with Eq.~(\ref{corr-gaussian}):
\be \label{h-h}
\la \lp h_{\r,t} - h_{\vec{0},0}\rp^2 \ra_0 \simeq 
{1\over 2\pi\kappa} \, 
\ln \lp {\r^2\over 4a^2} + {\eta\kappa |t| \over a^2}\rp \,.
\ee
At large $r$, $t$  
the correlation function (\ref{def-coscos}), as well as 
the current $\j^h$ correlation (\ref{fdt-wk}),
(\ref{def-deltasigma}), are thus of a power law form:
\be \label{coscos}
F(\r, t) 
= {1\over 2} \lp {a^2 \over \r^2/4 + \eta\kappa |t|}
\rp^{2\kappa^*/\kappa} \,. 
\ee
With $\kappa<\kappa^*$ in the correlated phase,
the function $F(\r, t)$ rapidly decays 
at large spatial and temporal separations.
We conclude from Eqs.~(\ref{fdt-wk}), (\ref{def-deltasigma}) and (\ref{coscos})
that in this case the current $\j^h(\r, t)$
space and time correlations 
have short memory and are local.
Hence one may treat $\j^h(\r, t)$ as an 
additional source of the Johnson-Nyquist noise, 
and our approach based on the fluctuation-dissipation
theorem (\ref{fdt-wk}) is \emph{a posteriori} justified.

Finally we consider the correction to the compressibility 
of the charged system due to the height fluctuations.
It can be obtained directly
from the statistical averaging with respect to the canonical distribution.
The compressibility $\nu$, defined as 
\be
\nu_\k^{-1} = \la \Phi_{-\k} \, \Phi_{\k} \ra %= U_{\k} 
\,, \quad  
{\rm with } \quad 
\Phi = {\delta \F \over \delta n} \,, 
\ee
acquires the perturbative correction 
as a result of the height-density coupling (\ref{F-int}).
The uniform ($k=0$) part $\delta\nu$ is obtained from 
\be \label{compressibility}
\lb\nu^{(0)} + \delta\nu \rb^{-1} = U_{\k=\vec{0}} + \delta U_{\k=\vec{0}}  \,,
\ee
with the height field-induced density-density interaction
\be \label{def-delta-nu}
\delta U_{\k} = 
\lambda^2 \int \! d^2\r \, e^{-i\k\r} \,
\la \cos {\pi h_{\r}\over b}\, \cos {\pi h_{\vec{0}}\over b}\ra_0 \,.
\ee
The correlator (\ref{def-delta-nu}) 
is given by Eq.~(\ref{coscos}) with $t=0$.
Thus the compressibility correction 
due to fluctuations
\be \label{nu-correction}
\delta \nu = -\lambda^2 
\lb U_{\k=0}\rb^{-2} 
\int\!d^2\r \, F(\r, t=0) \,.
\ee
This correction is {\it finite} in the correlated phase.
Indeed, the correlator in (\ref{def-delta-nu}) is local when 
$\kappa < 2\kappa^*$
which is consistent with the condition (\ref{kappa-compare}), 
and, thus, the integral in Eq.~(\ref{nu-correction})
is infrared-convergent.

%%%%%%%%%%%%%%%%%%%%%%%%%%%%%%%%%%%%%%%%%%%%%%%%%%%%%%%%%%%%%%%%%%%%%%%%%%%%%%%
\section{Summary}
\label{sec:discussion}

\nin
In the present work we analyze the problem of 
charge ordering and dynamics of classical
electrons on a 2d triangular array, by relating it to the better
studied problem of the triangular Ising antiferromagnet. 
The phase diagram (Fig.~\ref{phasediagram}) 
is found to be more rich than in the latter
problem due to long-range electron interactions
interplay with the geometrical frustration.
At low temperatures, the electron system freezes into
commensurate or disordered ground states in a continuous range of densities.
We demonstrate that transport can be employed to study charge ordering,
with the singularities of conductivity marking the phase transitions.

At intermediate temperatures, 
we idenitify the topological fluid phase
characterized by strong electron correlations.
The fluid state is described in terms of a nonlocal order parameter,
the height field. 
We find that the short range correlations in the charge system
correpsond to Gaussian fluctuations of the height field
in the presence of topological defects.
We determine numerically the effective stiffness of the height field,
and rule out the Berezinskii-Kosterlits-Thouless phase transition, concluding
that the topological defects remain unbound above freezing.
In this phase the electron 
transport, resulting from
short-range correlations enforcing
local continuity of the height field,
takes the form of ``free'' charge dynamics corresponding to 
height field fluctuations.
We found that both the topological defects and the 
height fluctuations contribute to transport, their relative contribution
depending on the concentration of the defects.

The ground state properties have been studied in this work 
for the simplest fractions $n=1/3$, $2/3$, and $n=1/2$.
In transport, freezing is manifest by singularities in zero bias conductivity which
drops to zero in the ordered phases. The sensitivity of 
transport to ordering of the electron system makes it a useful probe 
of 2d charge states.

While the stuation at $n=1/3$, $2/3$, $1/2$ 
is fairly conventional, the precise nature of 
ordering for generic density remains unclear.
Here several different scenarios can be anticipated.
One is a devil's staircase of incompressible states at any rational $n$.
In the relatively small system studied using MC dynamics, we have indeed 
found freezing into higher order fractions 
with relatively small denominator, such as 
the ``striped'' $n=3/7$ state shown in Fig.~\ref{snapshots} (D).
Similarly, other states with rational 
$n=p/q$ could become incompressible 
at low temerature, with $T_c$ decreasing with the increase of the 
denominator $q$, forming the devil's staircase.
Another possibility is the appearance,
besides the simple rational fractions, of a family of 
disordered ground states. One documented example of such a behavior is 
the preroughening phenomenon\cite{preroughening}
which describes a continuous phase transition into a disordered flat phase
of a 3d crystal surface, characterized by the height surface
with a disordered array of steps.
The tunability of the form of interaction in the quantum dot arrays
should allow to explore these, and other complex states.

%%%%%%%%%%%%%%%%%%%%%%%%%%%%%%%%%%%%%%%%%%%%%%%%%%%%%%%%%%%%%%%%%%%%%%%%%%
\section*{Acknowledgments}

\nin 
We are grateful to our colleagues Marc Kastner, Moungi Bawendi, and Nicole Morgan 
for drawing our attention to this problem and for useful discussions. 
We thank Sergei Korshunov for insightful comments on the manuscript,
and Pasquale Calabrese for a valuable discussion. 
The work at MIT was supported by the NSF MRSEC Award DMR 02-13282.
%L.L. and D.N. thank the Cavendish Laboratory, 
%where part of this work was carried out, for hospitality.
D.N. also acknowledges 
the NEC Fellowship for Advanced Materials (MIT), and support from
NSF MRSEC grant DMR 02-13706 (Princeton).

%%%%%%%%%%%%%%%%%%%%%%%%%%%%%%%%%%%%%%%%%%%%%%%%%%%%%%%%%%%%%%%%%%%%%%%%%%
\appendix

%%%%%%%%%%%%%%%%%%%%%%%%%%%%%%%%%%%%%%%%%%%%%%%%%%%%%%%%%%%
\section{Assigning height field to the array}

\nin
Here we describe the procedure by which 
the height values are assigned
to the sites of the array.
It should be noted that the height is well defined only
at $T=0$, in the absence of topological defects,
while at finite temperature it is defined only locally,
in the regions away from the defects.
As we found in Section \ref{sec:height-variable}, 
the defects are always present in the correlated
phase, since there is no defect binding phase transition.
This complication, however, is inessential,
since in the
number of defects is exponentially small at low enough temperature, 
and is often zero for 
the finite patch used in our simulation. 
We found that defining the height locally in a system 
with a small number of
defects does not lead to any significant errors
in the statistics
of fluctuations and, in particular,
does not affect the numerical value of the effective stiffness $\kappa$.

%%%%%%%%%%%%%%%%%%%%%%%%%%%%%%%%%%%%%%%%%%%%%%%%%%%%%%%%%%%%%%%%%%%%
\begin{figure}[t]
\includegraphics[width=3in]{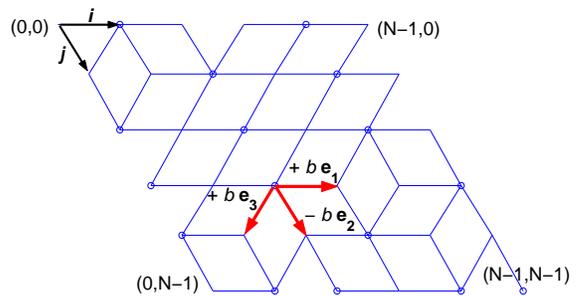} %coordinates.eps
\caption[]{(Color online)
Assigning auxiliary 3d coordinates to the sites of the array.
}
\label{fig:coordinates}
\end{figure}
%%%%%%%%%%%%%%%%%%%%%%%%%%%%%%%%%%%%%%%%%%%%%%%%%%%%%%%%%%%%%%%%%%%%

We work with an $N\times N$ rhombic patch of the 2d triangular
array, using the coordinate system aligned with the array,
as shown in Fig.\ref{fig:coordinates}.
The sites are labeled by integer coordinates 
$(i,j)$, $i, j = 0, ..., N-1$.
We place the origin in the upper left corner of the patch.
The first component, $i$,
is the site number counted along the horizontal axis (the patch upper edge). 
The numbers $i$ increase as we go from left to right.
The second component, $j$, is the site number along the axis which points downward
and to the right at the angle $\pi/3$ with the horizontal axis. 
The numbers $j$ increase as we go from the origin down and to the right, with
$j=N-1$ at the lower edge of the patch.
The conventional Cartesian coordinates $\r_m=(x_m, y_m)$ 
of the site $(i_m, j_m)$ are given by
\be \label{def-rm}
x_m = i_m\, a + j_m  \, {a\over 2}, \quad
y_m = -j_m\, a \, {\sqrt{3}\over 2} .
\ee
To calculate the height $h(\r_m)$ for each point $\r_m=(x_m, y_m)$ 
we first assign the auxiliary 3d coordinates 
\be \label{def-Rm}
\R(\r_m)=\sum_s R_s(\r_m) \, {\bf e}_s  \,.
\ee
Here the unit vectors ${\bf e}_s$, $s=1,2,3$, 
are the basis vectors of the auxiliary 3d simple cubic lattice.

To obtain the 3d coordinates $\R(i,j)$ for a particular charge configuration, 
we start from the upper left corner, $\R(i=0,j=0)\equiv \bf{0}$,
and assign the coordinates by the sequence of steps.
We shall call the sites $(i,j)$ and $(i',j')$  {\it connected},
if the charges at these sites are opposite, i.e., the bond between these sites
is not frustrated (and hence is drawn in Fig.~\ref{fig:coordinates}). 

To assign the height, we move from point to point in each row, 
$i=0,...,N-1$, with $j$ fixed,
repeating it for all $j=0,...,N-1$.
Suppose that the heights of 
all the sites with $i'\leq i$ and $j'\leq j$ are already defined.
The heights of the remaining one, two, or three sites that are connected to $(i,j)$ 
are defined by the following rules (Fig.~\ref{fig:coordinates}):

(i) If the site $(i+1,j)$ is connected to $(i,j)$, then $R_1(i+1,j)=R_1(i,j)+b$;

(ii) If the site $(i,j+1)$ is connected to $(i,j)$, then $R_2(i,j+1)=R_2(i,j)-b$;

(iii) If the site $(i-1,j+1)$ is connected to $(i,j)$, 
then $R_3(i-1,j+1)=R_3(i,j)+b$, with $b$ given by Eq.~(\ref{def-b}).

Once the 3d coordinates $\R(\r_m)$ are
defined, the height at each site $\r_m$ is obtained by 
projecting onto the (111) crystal direction:
\be \label{def-hi}
h(\r_m) = \sum_{s=1}^3 \, R_s (\r_m) \,.
\ee
We note that for
this procedure to work,
each site must be connected, in the above sense, to at least one neighbor.
Thus the height assignment may be impossible for charge configurations
with very low density of unfrustrated bonds. 
However, at low enough temperatures, in the correlated fluid phase,
at $n\simeq 1/2$, we do not encounter such configurations 
in our simulation.

Also, in the presence of defects, the above recipe assigns
heights unambiguosly, albeit in a somewhat \emph{ad hoc} way.
We do not explicitly exclude the configurations with defects from
the height statistics analysis. Instead, we estimate their relative 
contribution (Sec.~\ref{sec:cond-disl}) and find it to be small enough for 
the results (the stiffness value) to be affected.

%%%%%%%%%%%%%%%%%%%%%%%%%%%%%%%%%%%%%%%%%%%%%%%%%%%%%%%%%%%%%%%%%%%%%%%%%%

%\end{multicols}
\end{document}